\documentclass[12pt]{article}
\usepackage{float}

\usepackage{multicol}
\usepackage{mathtools}
\usepackage{listings}
\usepackage{color}
\usepackage{pdfpages}
\usepackage{amssymb}
\usepackage{tabularx}
\usepackage{bm}
\usepackage{url}
\usepackage{authblk}
  
\newcommand{\blind}{0}

\usepackage[round]{natbib}
\usepackage{amsmath,amssymb,amsthm,bm,enumerate,mathrsfs,mathtools}
\usepackage{latexsym,color,verbatim,multirow}
\usepackage[margin=1in]{geometry}
\usepackage{graphicx,hyperref}
\usepackage{caption}
\usepackage{footmisc}
\usepackage{lscape}
\usepackage{setspace}
\usepackage{newfloat}
\usepackage[para,online,flushleft]{threeparttable}
\DeclareFloatingEnvironment[name={Appendix Figure}]{suppfigure}
\usepackage{zwcommands}

\usepackage{titlesec}

\titleformat{\paragraph}
{\normalfont\normalsize\bfseries}{\theparagraph}{1em}{}
\titlespacing*{\paragraph}
{0pt}{3.25ex plus 1ex minus .2ex}{1.5ex plus .2ex}

\usepackage{xcolor}
\hypersetup{
  colorlinks,
  linkcolor={red!50!black},
  citecolor={blue!50!black},
  urlcolor={blue!80!black}
}

  \usepackage{circuitikz}

\usepackage{booktabs}

  \DeclarePairedDelimiter{\ceil}{\lceil}{\rceil}

\begin{document}
	\def\spacingset#1{\renewcommand{\baselinestretch}%
		{#1}\small\normalsize} \spacingset{1}


	\if0\blind
	{
		\title{\bf A Bayesian Approach to Restricted Latent Class Models for Scientifically-Structured Clustering of Multivariate Binary Outcomes}
		\author[1]{Zhenke Wu}
		\author[2]{Livia Casciola-Rosen}
		\author[2]{Antony Rosen}
		\author[3]{Scott L. Zeger}
		\affil[1]{Department of Biostatistics and Michigan Institute for Data Science, University of Michigan, Ann Arbor, MI 48109, USA; E-mail: {\tt zhenkewu@umich.edu}.}
		\affil[2]{Division of Rheumatology, Department of Medicine, Johns Hopkins University School of Medicine, Baltimore, Maryland, 21224, USA.}
		\affil[3]{Department of Biostatistics, Johns Hopkins University, Baltimore, MD 21205, USA.}
		\date{}
		\maketitle
	} \fi
	
	\if1\blind
	{
		\bigskip
		\bigskip
		\bigskip
		\begin{center}
			{\LARGE\bf A Bayesian Approach to Restricted Latent Class Models for Scientifically-Structured Clustering of Multivariate Binary Outcomes}
		\end{center}
		\medskip
	} \fi
	
\bigskip
\begin{abstract} 
In this paper, we propose a general framework for combining evidence of varying quality to estimate underlying binary latent variables in the presence of restrictions imposed to respect the scientific context. The resulting algorithms cluster the multivariate binary data in a manner partly guided by prior knowledge. The primary model assumptions are that 1) subjects belong to classes defined by unobserved binary states, such as the true presence or absence of pathogens in epidemiology, or of antibodies in medicine, or the ``ability" to correctly answer test questions in psychology, 2) a binary design matrix $\Gamma$ specifies relevant features in each class, and 3) measurements are independent given the latent class but can have different error rates. Conditions ensuring parameter identifiability from the likelihood function are discussed and inform the design of a novel posterior inference algorithm that simultaneously estimates the number of clusters, design matrix $\Gamma$, and model parameters. In finite samples and dimensions, we propose prior assumptions so that the posterior distribution of the number of clusters and the patterns of latent states tend to concentrate on smaller values and sparser patterns, respectively. The model readily extends to studies where some subjects' latent classes are known or important prior knowledge about differential measurement accuracy is available from external sources. The methods are illustrated with an analysis of protein data to detect clusters representing auto-antibody classes among scleroderma patients.
\end{abstract}
\noindent%
\vfill
\newpage
\spacingset{1.45} 

\section{Introduction}
\label{sec::intro}

Let $\mathbf{Y}$ be a $N\times L$ binary data matrix of $N$ observations with $L$ dimensions or features. Such multivariate binary data frequently arise as noisy measurements of presence or absence of a list of \textit{unobservable} or \textit{latent} binary variables $\bEta$ called \textit{states}. Suppose we seek to cluster such data subject to the hypothesis that a cluster is likely to be defined by individuals who share a relatively small number of states. That is, there exist subgroups of $\bEta$ vectors that take values on a relatively small number of elements in $\{0,1\}^M$ with $M\leq L$; let the subgroups be denoted by $\cA$. We propose a method for estimating \textit{scientifically-structured clusters} (SSC). Our method is most useful for a large dimension $L$ with an unknown number of clusters. Structured clustering for multivariate binary data has a number of potential advantages. If the underlying clusters differ from one another only at subsets of features, SSC can more accurately estimate these clusters than standard clustering methods such as latent class analysis and hierarchical clustering. SSC also results in more interpretable clusters. 

Consider three examples from medicine, psychology and epidemiology that motivate scientifically-structured clustering. Example 1 is to estimate subgroups of autoimmune disease patients using autoantibody data that have the potential to predict homogenous disease trajectories \citep[e.g.,][]{joseph2014association}. The observed binary responses are imperfect indicators of the presence or absence of specific autoantibody combinations detected in patient sera. Inherent limitation of the lab technique used to identify these autoantibodies (immunoprecipitation, IP) and biological and biochemical variability cause discrepancies between the expected presence/absence of each antibody and the observed values from IP assays. In addition, autoantigens (the specific proteins targeted by autoantibodies) frequently exist as multi-protein complexes, which we will refer to as ``machines" in this paper  \citep[e.g.,][also see Section \ref{sec::analysis}]{rosen2016autoantigens}. The medical goals are to define the ``machines" by their component proteins, and infer whether or not each patient has each machine, using the imprecise IP data.

The second example relates to cognitive diagnosis in psychological and educational assessment. The binary outcomes indicate a subject's responses to many diagnostic questions (``items"). The measurements reflect the person's long-term ``true" responses to these items, indicating a student's knowledge for correctly answering a test question absent guessing or other errors. These ``true" or ``ideal" responses are further assumed to define a smaller number of binary latent skills that indicate the presence or absence of the particular knowledge (called ``states" in the psychology literature). For example, teachers assess whether the student possesses basic arithmetic skills (e.g., addition, multiplication); and psychiatrists diagnose whether patients have certain mental disorders based on a subject's survey responses \citep[e.g.,][]{junker2001cognitive}. Each question or item is designed to measure a particular subset of latent states, where such item-latent-state correspondence may be known, partially known or unknown.

Example 3 is to estimate the causes of childhood pneumonia from a list of more than 30 different species of pathogens including viruses, bacteria and fungi \citep[e.g.,][]{obrien2017introduction}. The imperfect binary outcomes indicate whether or not each pathogen was detected by the polymerase chain reaction (PCR) or cell culture from two compartments: the nasopharyngeal (NP) cavity and blood. The binary latent states of scientific interest are the true presence or absence of the pathogens in a child's \textit{lung}, the site of infection that can seldom be directly observed in practice. This example differs from Example 1 in that the correspondence between each of the compartment-technology-pathogen diagnostic measurements (``features") and the latent lung infection (``state") is known because each measurement is designed to detect one specific pathogen and hence is expected to have higher positive rates in classes infected by that pathogen. In addition, the two measurements (NP with PCR and blood with cell culture) are known to have different error rates \citep[e.g.,][]{Hammitt2012, wu2016partially}. 

In each of these examples, the clustering of observations and subject-specific prediction of $\{\bEta_i\}$ comprise the scientific targets for inference. Our examples can be distinguished by:
\vspace{-0.0cm}
\begin{itemize}
\item[a)] whether the latent state variables ($\bEta_i$) are constrained or unconstrained to take values from a pre-specified subset $\cA$ where classes are defined by distinct values of $\bEta_i$, 
\item[b)] whether it is known, partially known, or unknown about the binary design matrix $\Gamma$ that specifies for each latent class the set of \textit{relevant} features having the highest positive response probability than other classes; and
\item[c)] the form of the conditional distribution of measurements given latent states and the design matrix ($\Gamma$) and response probabilities ($\Lambda$): $[\bY_i \mid \bEta_i, \Gamma, \Lambda]$.
\end{itemize}

This paper discusses a family of latent class models, referred to as restricted latent class models or RLCMs \citep[e.g.,][]{XuShang2017} specified by the three components listed above. The model formulation includes as special cases: probabilistic boolean matrix decomposition \citep{rukat2017bayesian}, subset clustering models \citep{hoff2005subset}, and partially latent class models \citep{wu2016partially} among others discussed in detail in Section \ref{sec::model_example}. The focus is on estimating clusters based on multivariate binary data that exhibit differential errors depending on the true latent class. The design matrix $\Gamma$ is assumed to be generated from a low-dimensional latent state vector $\bEta_i$. However, in many applications, the number of clusters and/or the set of latent states $(\cA)$ are not known \textit{in a priori} and must be inferred from data.

We discuss large-sample identifiability conditions for RLCM likelihood-based inference to motivate our posterior algorithm design. However, in finite samples, the likelihood function can be relatively flat before asymptotics concentrate the likelihood around the major mode. To improve finite-sample estimation efficiency at the expense of some bias, we specify sparsity-inducing priors that propagate into the posterior distribution to encourage few clusters with sparse latent state patterns. 

We begin this paper with a unified survey of restricted latent class models drawing on the previous work of \citet{wu2016partially}, \citet{wu2017nplcm}, \citet{hoff2005subset}, \citet{XuShang2017}. The second objective is to present  novel Markov chain Monte Carlo (MCMC) algorithms for Bayesian RLCMs with discrete component parameters building on the sampling techniques of \citet{jain2004split}, \citet{miller2017mixture} and \citet{chen2017bayesian}. Section \ref{sec::method} presents the model formulation including the likelihood, prior distribution and theoretical identifiability results. In Section \ref{sec::mcmc}, we present our MCMC algorithm to efficiently estimate posterior distributions for clusters. Section \ref{sec::simulation} compares via simulation the proposed clustering method to three common alternatives. Section \ref{sec::analysis} illustrates the methods with analysis of the autoantibody data for Example 1. The paper concludes with a discussion of model extensions and limitations. 

\section{Model}
\label{sec::method}

Let $\bm{Y}_i = (Y_{i1}, \ldots, Y_{iL})^\top \in\{0,1\}^L$ represent  a $L$-dimensional multivariate binary response for subject $i=1, \ldots, N$; Let $\mathbf{Y}$ collect data from all subjects. We assume each observation is associated with an unobserved or latent state vector $\bEta_i\in \cA$, where $\cA\subset \{0,1\}^M$ is a set of $M$-dimension binary vectors. 

Given a pre-specified dimension of latent states $M$, we first specify the likelihood $[\bY_i \mid \bEta_i, \Gamma, \Lambda]$ via restricted latent class models (RLCM) and then, among others, a prior distribution for $H=\{\bEta_i, i=1, \ldots, N\}$ that groups subjects by their binary patterns $\{\bEta_i\}$ (Supplementary Material A.1 extends the prior on $H$ to $M=\infty$). Let $\tilde{K}=|\cA|$ represent the number of groups with \textit{non-zero} population prevalence. Although it is no greater than $2^M$, $\tilde{K}$ can be unknown; And when $\tilde{K}$ is known and $\tilde{K} < 2^M$, $\cA$ can be unknown. The two steps jointly specify a so-called \textit{mixture of finite mixture model} for $\{\bY_i\}$ \citep{miller2017mixture}. In our setting, the salient feature of scientific import is the discrete component parameters $\{\bEta_i\} \subset \cA$ that requires additional handling in the posterior algorithm (Section \ref{sec::mcmc}). Section \ref{sec::model_example} discusses special cases of the RLCM relevant to the motivating examples. By taking $N$ to infinity in the likelihood, Section \ref{sec::identifiability} further studies theoretical limits of identifying unknown model parameters. 

\subsection{Latent Class Model}
For the traditional latent class model (LCM) \citep[e.g.,][]{Goodman1974}, we assume that the latent state vectors $\{\bEta_i\}$ take values from a set of binary patterns $\cA=\{\tilde{\bEta}_k, k=1, \ldots, \tilde{K}\}$,  where $\tilde{K} = |\cA|$ is the number of \textit{distinct} patterns. Latent classes differ in their latent state patterns. Given an observation's latent states $\bEta_i$, we assume the probability of observing a positive response of feature $\ell$ for subject $i$ is $\PP(Y_{i\ell}=1\mid \Lambda) = \lambda_{i\ell}$, $\ell =1, \ldots, L$, where $\Lambda = \{\lambda_{i\ell}\}$ is a $N\times L$ matrix of response probabilities; For $\Lambda$ and other matrices in this paper, we will use $\Lambda_{\star\ell}$ and $\Lambda_{i\star}$ to denote the $\ell$-th column and $i$-th row, respectively. A more useful, non-saturated model lets the response probability $\lambda_{i\ell}$ depend on the subject's latent state vector $\bEta_i$ via $\lambda_{i\ell} = \lambda_\ell(\bEta_i)$ where $\lambda_\ell: \cA \rightarrow [0,1]$. Because $\bEta_i$ can be one of $\tilde{K}$ elements in $\cA$, the classes have at most $\tilde{K}$ distinct response probabilities, referred to as \textit{between-class differential} measurement errors. 

The LCM has a \textit{conditional independence} assumption whereby the measurements from distinct dimensions are independent of one another given the latent class and response probabilities in that class, i.e. $Y_{i\ell} \perp Y_{i\ell'} \mid \bEta_i, \lambda_{\ell}(\cdot)$. Fitting LCM is to attribute, for example, a positive marginal association observed between two dimensions $\ell$ and $\ell'$ to their similar response probabilities that define the latent classes. Taken together, LCMs specify the conditional probability of observing a multivariate binary outcome $\bm{y}\in\{0,1\}^L$ by 
\begin{align}
\PP(\bm{Y}_i = \bm{y} \mid \bm{\eta}_i, {\lambda_\ell(\cdot)}) = \prod_{\ell=1}^L(\lambda_{i\ell})^{y_{i\ell}}(1-\lambda_{i\ell})^{1-y_{i\ell}}, \text{~where~} \lambda_{i\ell}=\lambda_{\ell}(\bEta_i).\label{eq::lcm_likelihood0}
\end{align}
Because $\bEta_i$ is not observed, it is integrated out of (\ref{eq::lcm_likelihood0}) with respect to its distribution $\PP(\bEta_i = \bEta \mid \bpi_{\tilde{K}}) = \pi_{\bEta} >0$, for $\bEta\in \cA$. Based on $N$ independent observations, the LCM likelihood takes the form of ``mixture of Bernoulli products": $\prod_{i=1}^N\sum_{\bEta \in \cA}\pi_{\bEta} \PP\left\{\bY_i \mid \bEta_i = \bEta, {\lambda_\ell(\bEta)}, \ell = 1, \ldots, L\right\} $. 

Given $\ell$, traditional LCMs impose no structure upon the response probability vectors except that they differ among classes almost surely: $\lambda_\ell(\bEta)\neq \lambda_\ell(\bEta')$ for latent classes $\bEta\neq \bEta'$. Let $\tilde{Z}_i \in \{1, \ldots, \tilde{K}\}$ indicate the \textit{unobserved} class assignment for observation $i$. An equivalent and more familiar formulation $\lambda_{i\ell} = \lambda_{\ell}(\tilde{Z}_i)$ results. The LCM approximates any multivariate discrete distribution for sufficiently large $\tilde{K}$ \citep[][Corollary 1]{dunson2009nonparametric} and, up to class relabeling, is generically identified whenever $L\geq 2\ceil{\tilde{K}}+1$ \citep[][Corollary 5]{allman2009identifiability}. Fitted LCM results will show the estimated response probability profiles that differ by class and can be interpreted as population heterogeneity in particular scientific contexts. Estimation of clusters in finite mixture models often makes use of $\{Z_i\}$, for example, by maximizing the plugged-in conditional posterior $\hat{Z}_i = \arg \max_{k=1, \ldots, \tilde{K}} \PP(Z_i = k \mid \mathbf{Y},\hat{\bpi}_{\tilde{K}})$ or a least-square estimate  of clusters based on distance from pairwise co-coclustering posterior probabilities $\hat{\pi}_{i,i'} = \PP(Z_i=Z_{i'}\mid \mathbf{Y})$ \citep{dahl2006model}.

\subsection{Motivation for Scientifically-Structured Classes}
The traditional LCM does not incorporate important prior scientific knowledge about how clusters (classes) structurally differ. In Example 1, autoimmune disease patients may differ in their antibody protein presence or absence patterns at $L=50$ protein landmarks over a grid of molecular weights. The focus is on estimating groups of patients who differ in their immune responses to unknown machines. We formulate this biological prior knowledge by introducing the following model parameters:
\begin{itemize}
\item[i)] An $M$ by $L$ machine matrix $Q$ where $Q_{m\ell} = 1$ indicates presence of landmark autoantigen protein $\ell$ in machine $m$; We refer to the rows in $Q$ as ``machine profiles". In addition, feature $\ell$ with $\sum_m Q_{m\ell}=0$ indicates landmark autoantigen protein $\ell$ is not targeted as part of any machine. 

\item[ii)] A patient-specific vector of length $M$ that represents the presence or absence of $M$ machines ($\bEta_i = (\eta_{i1}, \ldots, \eta_{iM})^\top$). For example, In Figure \ref{fig:likelihood_factorization}, for $M=3$, a subject with $\bEta_i = (1,0,1)^\top$ has Machines $1$ and $3$ (middle panel). The two machines produced her antibody proteins (left panel) subject to further errors. Given $Q$ and $\bEta_i$, we can represent the presence or absence of antibody proteins deterministically, for example, by $\Gamma_{i \star} = \bEta^\top_i  Q$ under a row-orthogonal $Q$ as illustrated in Figure \ref{fig:likelihood_factorization}. For feature $\ell$ with $\sum_{m}Q_{m\ell}=0$, we have $\Gamma_{i\ell}=0$ for all subjects.

\item[iii)] Positive rate parameters, the true- $(\btheta = \{\theta_{\ell} = \PP(Y_{i\ell}=1 \mid \Gamma_{i\ell}=1)\})$  and false- positive rates $(\bpsi = \{\psi_{\ell}=\PP(Y_{i\ell}=1 \mid \Gamma_{i\ell}=0)\})$. Two sources of stochastic variations contribute to the discrepancy between the expected presence of autoantibody ($\Gamma_{i\ell}=1)$ and the observed presence ($Y_{i\ell}=1$) or absence ($Y_{i\ell}=0$): selective immunological non-response to certain autoantigen proteins in a machine and experimental errors.  In a priori, we assume high true- and low false- positive rates ($\theta_\ell > \psi_\ell$) because GEA method is robust for detecting immunoprecipitated antibodies.

\end{itemize}

In summary, i) and ii) incorporate the prior knowledge that antibody proteins are produced in groups against autoantigen proteins coded by the rows of $Q$ and iii) is the measurement likelihood function that assigns probabilities to observed data accounting for stochastic variations. The other two examples in Section \ref{sec::intro} can be parameterized in the same way with known or unknown $Q$ (Section \ref{sec::model_example}).

Given subjects with $\Gamma_{i\ell}=1$ or $0$, the response probability $\lambda_{i\ell} = \theta_\ell$ or $\psi_\ell$ regardless of $i$'s class membership (e.g., true presence of antibody protein in serum no matter which machine it comes). Consequently, unlike traditional LCM, a new model where not all features exhibit difference in response probabilities $\{\lambda_{i\ell}, i = 1, \ldots, N\}$ is needed. Using separate class-specific estimates of $\{\lambda_{\ell}(1), \ldots,\lambda_{\ell}(\tilde{K})\}$ for features without actual between-class differential errors can be imprecise and will result in inferior clustering performance (Figure \ref{fig::intro_motivating_example}, d). RLCMs provide a general framework for specifying class response probability profiles to respect scientific structures through which we show achieve better clustering performance.

\begin{figure}[h]
\centering
\includegraphics[width=\textwidth]{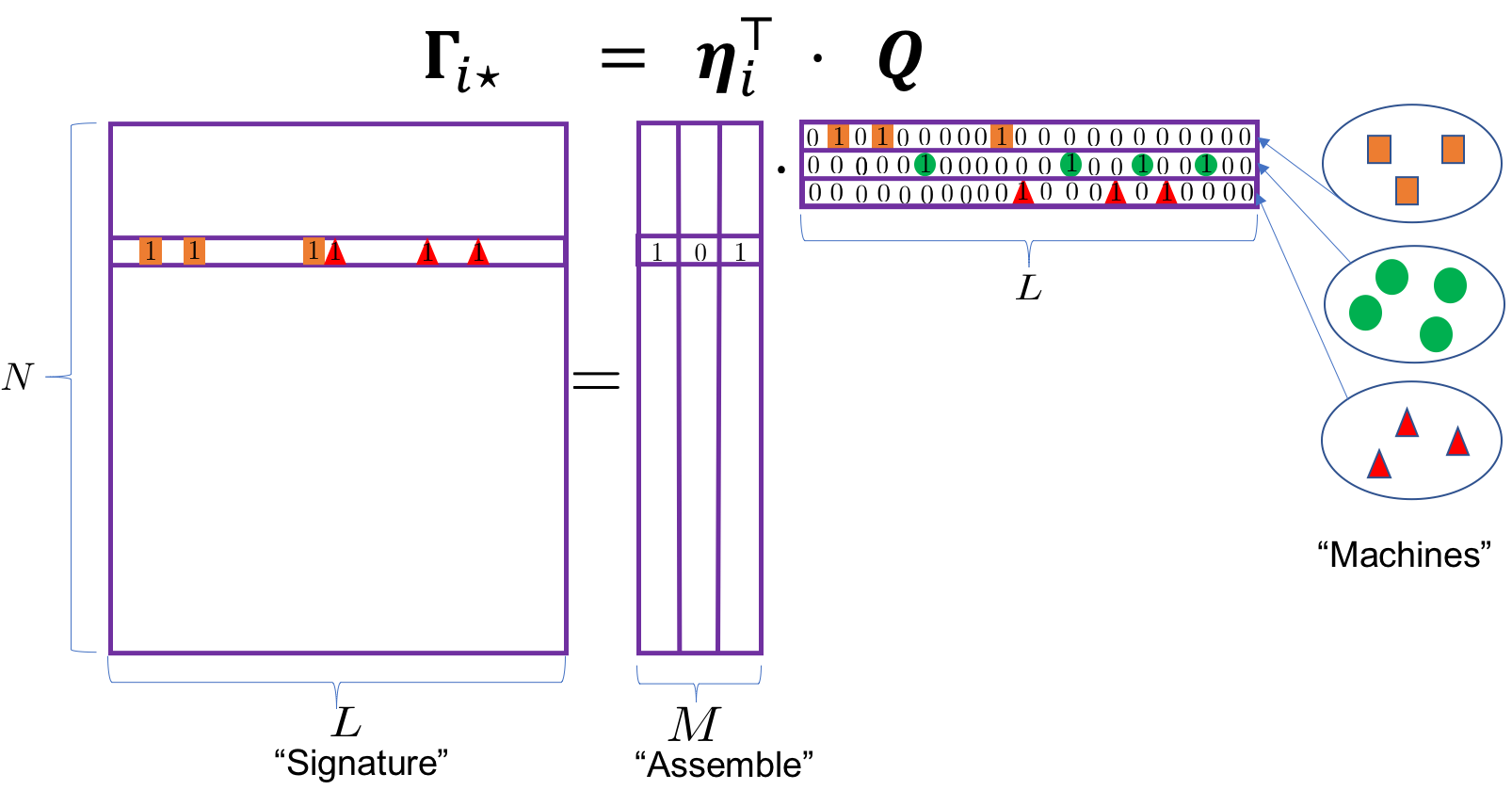}
\caption{Binary matrix factorization generates composite autoantibody signatures that are further subject to misclassification. The signature $\Gamma_{i\star}= \bm{\eta}_i^\top\bm{Q}$ assembles three \textit{orthogonal} machines with 3, 4 and 3 landmark proteins, respectively. The highlighted individual is expected to mount immune responses against antigens in Machines 1 and 3. See texts after model (\ref{eq::dino}).}
\label{fig:likelihood_factorization}
\end{figure}

\subsection{Restricted LCMs to Incorporate Scientifically-Structured Classes}
\label{sec::rlcm_formulation}
RLCMs assume equality among a subset of response probabilities across classes.  That is for some $\ell$, RLCMs assume $\lambda_{i\ell} = \lambda_{i'\ell}$ for some subjects in distinct latent classes ($\bEta_i \neq \bEta_{i'}$). The set of RLCM parameters therefore comprises a \textit{Lebesgue measure zero set} in the parameter space for the traditional unconstrained LCM. 

The restrictions on response probabilities in RLCMs are specified by introducing a binary design matrix $\Gamma= \{\Gamma_{\bEta,\ell}\} \in \{0,1\}^{\tilde{K}\times L}$ with latent classes and dimensions in the rows and columns, respectively. $\Gamma_{\bEta, \ell} =1$ represents a positive \textit{ideal} response for which subjects in latent class $\bEta$ will have the highest response probability at dimension $\ell$; $0$ for a negative \textit{ideal} response for which subjects in latent class $\bEta$ will have a lower response probability. If $\Gamma_{\bEta,\ell}=\Gamma_{\bEta',\ell}=1$ for two latent classes $\bEta$ and $\bEta'$, it is assumed that the $\ell$-th dimension is observed with identical positive response probabilities: $\lambda_{i\ell} = \lambda_{i'\ell}$. On the other hand, there can be more than one response probability if $\Gamma_{\bEta,\ell}=0$. That is, no equality constraint upon the response probabilities $(\lambda_{\bEta,\ell}=\lambda_{\bEta',\ell})$ is required for two latent classes with $\Gamma_{\bEta,\ell}=\Gamma_{\bEta',\ell}=0$.

In this paper, we focus on $Q$-restricted LCM where the design matrix $\Gamma$ is determined by the latent state vectors $\bEta$ and an $M$ by $L$ binary matrix $Q$, i.e., 
\begin{align}
\Gamma_{\bEta,\ell}  = \Gamma(\bEta, Q_{\star\ell}),\forall \bEta\in \cA, \ell = 1, \ldots, L, \label{eq::design_mat_by_Q}
\end{align} 
where the mapping or design matrix $\Gamma(\cdot, \cdot): \cA\times \{0,1\}^M \rightarrow \{0,1\}^{\tilde{K}\times L}$ needs to be specified in the context of the particular scientific study (e.g., $\Gamma_{i \ell} = \bEta_i^\top Q_{\star\ell}$ as in Figure \ref{fig:likelihood_factorization}). 

We now introduce scientific structures through the restriction of response probabilities. Let $\cA_{\ell} = \{\bEta \in \cA: \Gamma_{\bEta, \ell} = 1\}$ where $\cA_{\ell}$ collects latent classes with the highest response probability for dimension $\ell$ according to $\Gamma$ \citep{gu2018}. If $\cA_\ell \neq \emptyset$, we restrict the response probabilities at feature $\ell$ by
\begin{align}
\label{eq::rlcm_def}
\max_{\bEta \in \cA_\ell} \lambda_{\bEta, \ell} = \min_{\bEta \in \cA_\ell} \lambda_{\bEta, \ell} > \lambda_{\bEta', \ell}, \ell = 1, \ldots, L, \bEta' \notin \cA_\ell.
\end{align} 
Further, there can exist a class $\bEta \in \cA$ that gives rise to all-zero ideal responses $\Gamma_{\bEta\star}=\mathbf{0}_{1\times L}$.

To make the notation clear, in Example 1, $Y_{i\ell}$ represents the observed presence/absence of protein $\ell$ on the immunoprecipitation gel for patient $i$, $\bEta_i$ indicates this patient's latent class and which protein complexes (``machines") among the rows of $Q$ are present in patient $i$'s class, $Q_{m\star}, m=1,\ldots, M,$ indicates which proteins comprise Machine $m$,  and $\Gamma_{\bEta,\ell}$ indicates whether or not protein $\ell$ is present from any machines in latent class $\bEta$. The class with no machine has all zeros in its row of $\Gamma$. The probability of observing a protein given it is present is its true positive rate or sensitivity. The probability of observing the protein given it is absent is its false positive rate or one minus its specificity. The sensitivity for a given protein is assumed to be the same regardless from which machine(s) it comes; the specificities are allowed to vary across proteins. Finally, the true positive rates are assumed to be larger than the false positive rates.

Finally, we specify the measurement likelihood through parameterization of the response probabilities
\vspace{-1cm}
\begin{align}
\lambda_{i\ell} = \lambda^R_\ell\left(\bEta_{i};\bbeta_\ell, Q_{\star\ell} \right) \in [0,1], \label{eq::lambdaR}
\end{align}
\vspace{-1cm}

\noindent where $\lambda^R_\ell$ specifies the response probability at feature $\ell$ with ``restriction" (\ref{eq::rlcm_def}). The restriction is prescribed by $\Gamma_{\bEta_i, \ell}$ which is further determined by $\bEta_i$ and $Q_{\star\ell}$ in (\ref{eq::design_mat_by_Q}). $\lambda^R_\ell$ also depends on $\bEta_i$ and unknown real-value parameters $\bbeta_\ell$ according to particular parametric models; See model (\ref{eq::logit_cdm}) below for an example. In what follows, we use $\Gamma_{i\ell}$ to denote $\Gamma_{\bEta_i, \ell}$ unless otherwise noted.

Motivated by our applications, we present an equivalent formulation for $\lambda^R_{\ell}$ that separate true and false positive rates. Let $K^+_\ell = \#\{\lambda_{i\ell}: \bEta_i \in \cA_\ell\}$ ($K^-_\ell = \#\{\lambda_{i\ell}: \bEta_i \notin \cA_\ell\}$) be the number of distinct response probability levels at feature $\ell=1, \ldots, L$. In RLCMs, we have $K^+_\ell = 1$ and $K^-_\ell \geq 1$, $\ell = 1, \ldots, L$ ( see Table S1 in Supplementary Materials that tabulate the number of distinct response probabilities at dimension $\ell$, $(K^+_\ell, K^-_\ell)$, for other variants of LCMs). Let $\theta_l$ be the maximum response probability at feature $\ell$ and $\bm{\psi}_\ell = \{\psi_{l1}, \ldots, \psi_{l,K^-_\ell}\}$ be the rest of response probabilities, respectively. Given $\bEta_i = \bEta \notin  \cA_\ell$,  let $v_i = v(\bEta_i, \ell)$, where $v(\cdot, \cdot)$: $(\bEta_i,\ell) \mapsto v$ is the integer-valued function that selects among $\bm{\psi}_\ell$ her associated response probability $\psi_{\ell, v_i}$ at feature $\ell$. The parameters $\theta_\ell$ and $\bpsi_\ell$ may be further parameterized by $(\bbeta_\ell,Q_{\star\ell})$ as in (\ref{eq::lambdaR}). For models with $K^-_\ell =1$, $v(\cdot)=1$; Otherwise, $\nu(\cdot,\cdot)$ depends on $\cA$ (the set of possible pattern of $\bEta_i$), the specific functional form of $\lambda^R_\ell(\cdot)$ and parameter values of $(\bbeta_\ell, Q_{\star\ell})$ in a RLCM  (see the example (\ref{eq::logit_cdm}) in Section \ref{sec::model_example}; The traditional LCM results by setting $Q=1_{M\times L}$ and under $K^++K^-=\tilde{K}$ for each $\ell$). 

 In this paper, because we focus on models with the structure in (\ref{eq::design_mat_by_Q}), we can equivalently represent the response probability parameters $\lambda^R_\ell$ in (\ref{eq::lambdaR}) by 
\begin{align}
 \lambda^R_\ell(\bEta_i; \bbeta_\ell, Q_{\star\ell}) =\left\{\theta_\ell\right\}^{\Gamma_{\bEta_i,\ell}}\cdot \left\{\bpsi_{\ell,v(\bEta_i,\ell)}\right\}^{1-\Gamma_{\bEta_i,\ell}} \in [0,1],\label{eq::response_prob}
\end{align}
where $\bbeta_\ell = \{\btheta =\left\{\theta_\ell\}, \Psi = \{ \bpsi_\ell\}\right\}$ with constraints $\theta_\ell > \psi_{\ell, v}, \forall v=1, \ldots, K^-_\ell$. RLCMs therefore let observations with latent state patterns in $\cA_\ell$  take identical and the highest probability than other classes in $\cA^c_\ell$. Other classes in $\cA_\ell^c$ respond with lower probabilities at dimension $\ell$. We now discuss some examples.

\subsection{Examples of RLCMs in the Literature}
\label{sec::model_example}

Special cases of restricted LCMs result when $K^+_\ell=K^-_\ell=1$. For example, a class of models assumes the response probabilities 

\vspace{-1cm}
\begin{align}
\lambda_{i\ell} & =  \theta_{\ell}^{\Gamma_{i\ell}}(\psi_{\ell})^{1-\Gamma_{i\ell}}, ~~~~\Gamma_{i\ell}= 1-\prod_{m=1}^M(1-\eta_{im})^{Q_{m\ell}}. \label{eq::dino}
\end{align} 
\vspace{-1cm}

\noindent Consider $N$ subjects each responding to $L$ items where $Q_{m\ell}=1$ means item $\ell$ requires positive latent state $m$, otherwise $Q_{m\ell}=0$. This model, referred to as partially latent class models in disease epidemiology \citep[][PLCM]{wu2016partially} or Deterministic In and Noisy Or (DINO) in cognitive diagnostic models \citep[e.g.,][DINO]{templin2006measurement} that needs just one required state ($\{m: Q_{m\ell}=1\}$) for a positive ideal response $\Gamma_{i\ell}=1$. Imposing constant and symmetric error rates $\theta_{\ell}=\psi_{\ell}$, $\ell = 1, \ldots, L$, the one-layer model of \citet{rukat2017bayesian} results. The model can also be viewed as Boolean matrix factorization \citep[BMF,][]{miettinen2008discrete} by noting that $\Gamma_{i\ell}=  \lor_{m=1}^M \eta_{im}Q_{m\ell}$ where the logical ``OR" operator $``\vee"$  outputs one if any argument equals one. The rows in ${Q}$ are basis patterns for compactly encoding the $L$ dimensional $\Gamma_{i\star}$ vector by $M(\ll L)$ bits. BMF further reduces to nonnegative matrix factorization \citep[e.g.,][]{lee1999learning} $\Gamma=HQ$ where $H=\{\eta_{im}\}$ if ${Q}$ has orthogonal rows.  See Supplementary Materials A.2 for a connection to subset clustering in \citet{hoff2005subset}. A second two-parameter example results by assuming  $\Gamma_{i\ell}= \prod_{m=1}^M(\eta_{im})^{Q_{m\ell}}$ \citep[e.g.,][]{junker2001cognitive}. This model, referred to as Deterministic In and Noise And (DINA) gate model in the cognitive diagnostic literature, assumes a conjunctive (noncompensatory) relationship among latent states $m=1, \ldots, M$. That is, it is necessary to possess \textit{all} the attributes (states) indicated by non-zero elements in $Q_{\star\ell}$ to be capable of providing a positive ideal response $\Gamma_{i\ell}=1$. The model also imposes the assumption that possessing additional unnecessary attributes does not compensate for the lack of the necessary ones. These two-parameter models are equivalent upon defining $\eta_{im}^*=1-\eta_{im}$, $\Gamma_{i\ell}^*=1-\Gamma_{i\ell}$, $\psi^*_{\ell}=1-\psi_{\ell}$ and $\psi^*_{\ell}=1-\theta_{\ell}$ \citep{chen2015statistical}. There are several other examples in this category as discussed by \citet{xu2017identifiability}.

Two-parameter models assume that ``$\Gamma_{\bEta,\ell}=\Gamma_{\bEta',\ell}=0$ implies identical response probabilities $\lambda_{\bEta,\ell} = \lambda_{\bEta',\ell} = \psi_{\ell}$", regardless of the distinct patterns $\bEta \neq \bEta'$. In practice, deviation from such assumptions occurs if $\bEta$ has more nonzero elements than $\bEta'$ and alters the response probabilities, i.e., $K^-_\ell>1$. Multi-parameter models where $K^-_\ell >K^+_\ell=1$, popular in multidimensional item response theory, is readily specified for example by assuming an all-effect model: $\lambda_{i\ell} = \lambda^R_{\ell}(\bEta_i; \bm{\beta}_\ell, Q_{\star\ell})= \textrm{expit}\left\{\bm{\beta}_\ell ^\top \bh(\bEta_i, Q_{\star\ell})\right\} = $

{\footnotesize
\vspace{-1cm}
\begin{align}
 & \textrm{expit} \left\{ \beta_{\ell0}+\sum_{m=1}^M\beta_{\ell m}(Q_{m\ell}\eta_{im})+\sum_{m < m' }\beta_{\ell m m'}(Q_{m\ell}\eta_{im})(Q_{m'\ell}\eta_{im'})+\ldots+\beta_{\ell 12\ldots M}\prod_{m}(Q_{m\ell}\eta_{im})\right\} \label{eq::logit_cdm}
\end{align} 
}
\vspace{-1cm}

\noindent that includes higher order interactions among latent states required by an item \citep{henson2009defining}; Here $\text{ expit}(x)=\frac{\exp(x)}{1+\exp(x)}$. When $\prod_{m=m_1, \ldots, m_s}Q_{m\ell}=0$, this saturated model needs no $\beta_{\ell,m_1\ldots m_s}$ term. Setting second or higher order terms to zero, an additive main-effect model results. The effects of latent states need not be additive. For example, \({\sf log}(\lambda_{i\ell}) = \beta_{\ell 0}+\sum_{m=1}^M \beta_{\ell m}Q_{m\ell}\eta_{im}\) specifies a multiplicative model that penalizes the absence of an required latent state $m$ if $Q_{m\ell}=1$. 

Table S1 in Supplementary Materials summarizes these and other variants of LCMs by specifications of the latent state space, design matrix, and measurement processs.

\subsection{Identifiability}
\label{sec::identifiability}

There are two sources of indeterminancy in restricted LCMs: invariance of the likelihood function to permutation of the ordering of the latent states and over-parameterized models. The permutation invariance manifests itself as a multimodal posterior distribution. Where $Q$ is unknown, we address the permutation invariance by labeling the latent states, one dimension at a time, by the non-zero patterns of the corresponding rows in an estimated $Q$.  We address the over-parameterization by introducing prior distributions that encourage \textit{in a priori} few clusters hence a small number of parameters via mixture of finite mixture models \citep{miller2017mixture}. It helps to show identifiability results or lack thereof to motivate such sparsity-inducing priors.

Given $\tilde{K}$ and $M$, identifiability conditions characterize the theoretical limits of recovering the unknown model parameters ($Q$, $\Lambda$, $\bpi_{\tilde{K}}$) from the likelihood for all or a subset of the parameter space. We first discuss the identifiability of $Q$ because it is needed for interpreting latent states (see Section \ref{sec::analysis}) and for estimating both $H$ and $\bpi_{\tilde{K}}$. Based on the likelihood $[\bY_i \mid \bpi_{\tilde{K}}, \Lambda, \Gamma=\Gamma(Q)]$ with a given $Q$ and a saturated $\cA$ (or ``full diversity": $\pi_{\bEta}>0, \forall \bEta \in \cA=\{0,1\}^M$), \citet{xu2017identifiability} studied sufficient conditions for \textit{strict} identifiability of $\Lambda$ and $\bpi_{\tilde{K}}$ over the entire parameter space in RLCMs. Under weaker conditions upon the design matrix $\Gamma$ (instead of $Q$) and possibly non-saturated $\cA$, \citet{gu2018} established conditions that guarantee \textit{partial} identifiability for general RLCMs which means the likelihood function is flat over a subset of the parameter space. When $Q$-matrix is completely unknown, it is possible to identify $\{\bpi_{\tilde{K}}, \Lambda, Q\}$ just using likelihood $[\bY_i \mid \bpi_{\tilde{K}}, \Lambda, \Gamma=\Gamma(Q)]$. In particular, \citet{chen2015statistical} provided sufficient conditions for the special cases of DINA and DINO models (see Section \ref{sec::model_example}); \citet{XuShang2017} further generalized them to general RLCM: ($Q$, $\Lambda$, $\bpi_{\tilde{K}}$) are strictly identifiable (up to row reordering of $Q$) in RLCMs with saturated $\cA$ if the following two conditions hold: 
\begin{itemize}
\item[C1)] The true ${Q}$ can be written as a block matrix ${Q}=[{I}_M; {I}_M; \tilde{{Q}}]$ after necessary column and row reordering, where $\tilde{{Q}}$ is a $M\times (L-2M)$ binary matrix and
\item[C2)] $(\Lambda_{\bEta,\ell}, \ell > 2M)^\top \neq(\Lambda_{\bEta',\ell}, \ell > 2M)^\top$ for any $\bEta \neq \bEta'$ and $\bEta\succeq \bEta'$,
\end{itemize}
where $\bm{a}\succeq \bm{b}$ for $\bm{a} = \{a_j\}$ and $\bm{b}=\{b_j\}$ if and only if $a_j\geq b_j$ holds element-wise.

Because condition (C2) depends on $Q$, $\Lambda$ and row and column permutations, the number of operations to check (C2) increases exponentially with $M$, $\cO((L-2M)2^MM)$, for a saturated $\cA$ with $2^M$ patterns of latent state vectors. We instead use condition (C3) that just depends on $Q$ and that is invariant to row or column permutations:
\begin{itemize}
\item[C3)] Each latent state is associated to at least three items,  $\sum_{\ell=1}^L Q_{m\ell} \geq 3$ for all $m$.
\end{itemize} 
Condition (C3) enables convenient restrictions in MCMC sampling and takes just $\cO(LM)$ operations to check. For special cases of RLCM, the DINA and DINO models (Section \ref{sec::model_example}) with a saturated $\cA$, Conditions (C1) and (C3) suffice to identify ($Q$, $\Lambda$, $\bpi_{\tilde{K}}$) \citep[Theorem 2.3,][]{chen2015statistical}. 

Posterior algorithms typically restrict MCMC sampling of non-identified parameters by identifiability conditions to prevent aggregation of posterior probability mass from multiple modes. For example, in factor analysis of multivariate continuous data, one can restrict the loading matrices in lower triangular forms \citep[e.g.,][]{geweke1996measuring}.  Alternatively, one may first perform MCMC sampling with weak and simple-to-check constraints without fully ensuring identifiability and just check afterwards whether the parameters are conditionally identifiable. One then performs necessary deterministic transformations on parameters that may only be identified up to equivalent classes to pick coherent and economical representatives, for example, by relabeling sampled mixture components at each iteration or varimax rotations of factor loading matrices in classical Gaussian factor analysis \citep[e.g.,][]{rovckova2016fast}. 

We initialize the sampling chain from the set defined by simple identifiability conditions (C1) and (C3) and only check afterwards at each iteration whether the parameters are conditionally identifiable according to conditions (C1) and (C2) that are stronger and computationally more expensive. The relabeling of the latent states is done by inspecting the non-zero patterns in the rows of $Q$ (Step 7, Supplementary Material C.1).

In applications where $Q$ is unknown with $M<L/2$, we focus on the set of $Q$-matrices that satisfy both (C1) and (C3):
\begin{equation}
\mathcal{Q} = \{{Q} \in \{0,1\}^{M\times L}: {Q}=P_1{Q}^\dagger P_2, ~{Q}^\dagger=[{I}_M; {I}_M; \tilde{{Q}}], ~\tilde{{Q}}\mathbf{1}_{L-2M}\succeq\mathbf{1}_{L-2M}\},\label{eq::identifying_constraint}
\end{equation}
where ${P}_1$ and $P_2$ are $M$- and $L$-dimensional permutation matrices for rows and columns, respectively. The constraint $\mathcal{Q}$ also greatly facilitates posterior sampling by focusing  on a small subset of binary matrices. In fact, among all $M$ by $L$ binary matrices, the fraction of $Q\in\mathcal{Q}$ is at most $\frac{{L \choose 2M}\left[2^{(L-2M)M}\right]}{2^{L\cdot M}}$ and quickly decay as the number of machines $M$ increases. In some applications it may also simplify posterior inference by exploiting further assumptions upon $Q$ for example partially known $Q$ or non-overlapping (i.e., orthogonal) rows of $Q$.  See Supplementary Materials A.3 and A.4 for other identifiability considerations that motivate our posterior algorithms.

\subsection{Priors}
\label{sec::prior}
Given $M$, we specify the prior for $H=\{\bEta_i\}$ with cluster structure among $N$ subjects in five steps: 1) Generate the vector of probabilities of a subject $i$ belonging to each of $K$ clusters $\bpi_K=(\pi_1, \ldots, \pi_K)^\top$ where $K$ is possibly unknown and sampled from its prior $p_K(\cdot)$; 2) Partition observations by indicators $Z_i\overset{i.i.d}{\sim} {\sf Categorical}(\bpi_K)$; Suppose we obtain $T$ distinct $\{Z_i\}$ values; 3) Draw the vector of marginal probabilities of each latent state being active $\bp = \{p_m\}$;  4) Draw from $[\bEta^*_j \mid \bp,M]$, for clusters labeled  $j=1, \ldots, T$, where ``$^*$" indicates cluster-specific quantities; 5) Combine $\{\bEta^*_j\}$ and $\{Z_i\}$ to obtain subject-specific latent states $\bEta_i = \bEta^*_{Z_i}$, $i=1, \ldots, N$.

\subsubsection{Prior for Partitioning Observations}
\label{sec::clustering}

Though used interchangeably by many authors, we first make a distinction between a ``component" that represents one of the true mixture components in the specification of a mixture model and a ``cluster" that represents one element in any partition of observations. Let $K$ be the number of mixture components in the \textit{population} and $T$ the number of clusters in the \textit{sample} \citep{miller2017mixture}. 

To establish notation, let $Z_i \in \{1, 2, \ldots, K\}$ be the subject-specific component indicators, 
$E_z= \{i: Z_i =z\}$ the set of subjects in component $j$, $\mathcal{C} = \{C_j: |C_j|>0\}$ the partition of $N$ subjects induced by $\bZ=\{Z_i, i=1, \ldots, N\}$; Note the partition $\cC$ is invariant to component relabeling. Let $T =|\mathcal{C}|$ be the number of clusters formed by the $N$ \textit{subjects}; it may differ from $K$, the number of components for the \textit{population}. Further let $C \in \mathcal{C}$ denote one of the clusters in partition $\mathcal{C}$; let $j$ be the index associated with cluster $C_j$, for $j \in\{1, \ldots, T\}$. Let $\mathcal{C}_{-i} = \{C_j \setminus \{i\}: |C_j \setminus \{i\}|>0\}$ be the partition of subjects excluding subject $i$. For simplicity, let $\mathbf{Y}_{C} = \{\bm{Y}_{i}, i\in C\}$ be the collection of data in a cluster $C\in \mathcal{C}$. Finally, let $\bEta_i$ be the latent state vector for subject $i=1, \ldots, N$, and $\bEta^*_j$ be the latent state vectors for cluster $j=1, \ldots, T$.
 
We assume the indicators $\bZ$ are drawn as follows:

\vspace{-1cm}
\begin{align}
 {\sf Number~of~components:~~} K  & \sim p_K, \label{eq::mixture1}\\
{\sf Mixing~weights:~~} \bpi_K & \sim {\sf Dirichlet}(\gamma, \ldots, \gamma),\label{eq::mixture2}\\
{\sf Cluster~indicators:~~} Z_i & \sim {\sf Categorical}\{\bpi_K=(\pi_1, \ldots, \pi_{K})\}, i = 1, \ldots, N,\label{eq::mixture3}
\end{align}
\vspace{-1cm}

\noindent where $p_K$ is a probability mass function over  non-zero integers $\{1,2, \ldots\}$ and $\gamma >0$ is the hyperparameter for symmetric $K$-dimensional Dirichlet distribution. Note that though $\tilde{K}\leq 2^M$, $K$ is not upper bounded (unless constrained through the support of $p_K$). The prior of partition $\mathcal{C}$ induced by (\ref{eq::mixture1}-\ref{eq::mixture3}) is
\(
p(\mathcal{C}\mid \gamma, p_K(\cdot)) = V_N(T)\prod_{C\in \mathcal{C}} \gamma^{(|C|)}, \label{eq::partition_prior}
\)
where $V_N(T) = \sum_{k=1}^\infty \frac{k_{(T)}}{(\gamma k)^{(N)}}p_K(k)$, $T=|\mathcal{C}|$ is the number of blocks/partitions for $N$ subjects and by convention $k^{(n)} = k\cdot (k+1)\cdots (k+n-1)$, $k_{(n)} = k\cdot (k-1)\cdots (k-n+1)$, and $k^{(0)}=k_{(0)}=1$, $k_{(n)}=0$ if $k<n$ \citep[][]{miller2017mixture}.

\subsubsection{Prior for $H^*$}
\label{sec::prior_H}
Given $\{Z_i\}$, we draw the latent state vector $\bEta^*_j\in \{0,1\}^M$ for which $Z_i=j$ indicates, referred to as ``component-specific parameters" in mixture models. We discuss priors for these \textit{discrete} component parameters according as $\cA$ is known or not.

\label{sec::unknown_M}
\noindent \underline{\it Pre-specified $\cA$.}  In applications such as Example 3, pre-specifying $\cA$ is appealing when the scientific interest lies in itemized characterization of the population fractions for each element of $\cA$.  Given $\cA$, the cluster membership indicators $\{Z_i\}$ take value from $\{1, \ldots, T\}$ where $T= \tilde{K}$. Existing approaches then assign to each cluster one of $\{\bEta^*_1, \ldots, \bEta^*_T\}$ by enumerating the distinct \textit{known} elements in $\cA$. For example, see \citet{chen2015statistical} for $\cA=\{0,1\}^M$, $\tilde{K}=2^M$.  \citet{wu2016partially} analyzed data from Example 3 and specified $\cA = \{\be_1, \ldots, \be_M,\mathbf{0}_M\}$ among pneumonia \textit{cases} that represents latent states as the lung infection caused by pathogen $1, 2, \ldots, M$ or none-of-the-above and $\bEta_i=\mathbf{0}_M$ among observed \textit{controls}. Absent the uncertainty in $\cA$, simpler posterior sampling algorithms result. 

In practice, to avoid misleading estimates based on a pre-specified $\cA$ subject to potential misspecification, analysts may conservatively specify $\cA=\{0,1\}^M$. However, $\bEta_i=\bEta^*_{Z_i}$ then take its value from a space that grows exponentially with $M$ (e.g., $M=30$ in Example 3). Consequently, upon fitting the model for inferring $\pi_{k}, k=1,\ldots, \tilde{K}(=2^M)$, although many elements in $\cA$ may receive low posterior probabilities, none is exactly zero. Important elements in $\cA$ are commonly selected by \textit{ad hoc} thresholding. In addition, pre-specifying $\cA\subsetneqq \{0,1\}^M$ does not address the question of what are the distinct latent state patterns $\tilde{\bEta}^*_j$ in the data. 

\noindent \underline{\it  Unknown $\cA$}. Absent knowledge of $\cA$, we draw\textit{ in a priori} the component-specific parameters $H^*=\{\eta_{jm}^*\}$ in two steps for regularizing $\bEta^*_j$ towards sparsity:

\vspace{-1cm}
\begin{align}
{\sf probability~of~an~active~state:~~} p_{m} \mid \alpha_1, \alpha_2 & \sim {\sf Beta}(\alpha_1\alpha_2/M, \alpha_2), \label{eq::finite_ibp_step1}\\
{\sf latent~states:~~}  \eta^*_{jm} \mid p_m & \sim {\sf Bernoulli}(p_m), j=1, \ldots, T,\label{eq::finite_ibp_step2}
\end{align}
\vspace{-1cm}

\noindent for $m=1, \ldots, M$. Note that it is possible that $\bEta^*_j = \bEta^*_{j'}$ for some $j,j'=1, \ldots, T$ where equality holds element-wise. For example, $\bEta^*_{Z_i}$ may equal  $\bEta^*_{Z_i'}$ even if $Z_i\neq Z_{i'}$. Because we are interested in estimating distinct $\bEta^*_j$'s that represent distinct values of scientific latent constructs, we will merge such clusters $j$ and $j'$ into one, referred to as a ``scientific cluster"; We denote it by $\tilde{\cC}$. We also denote the unique values in $H^* = \{\bEta^*_j, j=1, \ldots, T\}$ by $\tilde{H}^{*}=\{\tilde{\bEta}^*_j, j=1, \ldots, \tilde{T}\}$. Supplementary Material A.5 and A.6 further remarks on the induced priors on the partitions $\cC$ and $\tilde{\cC}$.

\begin{remark}
\label{remark::pseudo_cluster}
The $K$ introduced in the prior specification is to make it not upper bounded and therefore differs from $\tilde{K}$. The latter represents the number of \textbf{distinct} latent state vectors in the population and must be no greater than $2^M$. $\tilde{\bEta}_k, k= 1, \ldots, \tilde{K}$ represent the set of true distinct latent state vectors in the population; while $\bEta^*_j, j=1, \ldots, T$ ($T\leq K$) represent the realized latent state vectors that are \textbf{possibly duplicated} in the data generating process (\ref{eq::finite_ibp_step2}) or the posterior sampling. With unconstrained $K$, we are able to build on the algorithm of \citet{miller2017mixture} that does not bound the number of mixture components. The resulting algorithm works for general mixture of finite mixture models with \textbf{discrete} component distributions (Section \ref{sec::mcmc}).  
\end{remark}

 By Beta-Bernoulli conjugacy, we integrate $[H^* \mid \bp][\bp \mid \alpha_1, \alpha_2]$ over $\bp$ to obtain the marginal prior: 
\begin{align}
pr(H^*) = \prod_{m=1}^M \frac{(\alpha_1\alpha_2/M)\Gamma(s_m +\alpha_1\alpha_2/M)\Gamma(T-s_m+\alpha_2)}{\Gamma(T+\alpha_2+\alpha_1/M)},\label{eq::finite_ibp}
\end{align}
where $\Gamma(\bullet)$ is the Gamma function and $s_m = \sum_{m=1}^T \eta^*_{jm}$, $j=1, \ldots, M$. Holding $\alpha_2$ constant, the average number of positives among $\bEta^*_j$ decreases with $\alpha_1$; Holding $\alpha_1$ constant, the latent state vectors, $\bEta^*_j$ and $\bEta^*_{j'}$, $j\neq j'$, become increasingly similar as $\alpha_2$ decreases. In fact, the probability of two subjects with distinct cluster indicators $Z_i$ and $Z_{i'}$ have identical $m$-th latent state, $\PP[\eta^*_{im} = \eta^*_{i'm} \mid Z_i = j, Z_{i'}=j', j\neq j',  \alpha_1, \alpha_2] = \EE\{p_{m}^2+(1-p_m)^2 \mid \alpha_1,\alpha_2\} = 1-2\frac{\alpha_1}{\alpha_1+M}\left(1-\frac{\alpha_1\alpha_2+M}{\alpha_1\alpha_2+\alpha_2M+M}\right)$ approaches one when $\alpha_2$ goes to zero. In what follows, $\alpha_2$ is set to $1$ which offers good clustering results in simulations and data analyses. Finally in applications where no pooling across $j$ is needed, one can set $p_m=0.5$ to specify uniform distribution over all possible patterns over $\cA=\{0,1\}^M$.

\subsubsection{Priors for Other Model Parameters}
\label{sec::prior_other}

We focus on the situation where $Q$ is completely unknown. Let $Q$ be uniformly distributed over the constrained space in $\{0,1\}^{M\times L}$ defined by  (\ref{eq::identifying_constraint}). In applications where $Q$ is not fully identifiable and/or encouraged to be different among its rows in finite samples, we specify sparsity priors for each column of $Q$ to encourage proteins to be specific to a small number of machines (see Supplementary Material A.6). 

 We specify the priors for response probabilities $\Lambda=\{\lambda_{i\ell}\}$ in (\ref{eq::response_prob})  to satisfy the monotonic constraints in (\ref{eq::rlcm_def}) as follows
\begin{align}
\psi_{\ell,v}  & {\sim} {\sf Beta}(N_{\psi} a_\psi, N_{\psi}(1-a_\psi)),  v = 1, \ldots, K^-_\ell, \text{~constrained to~}\Delta = \left\{\{\bpsi_{\ell}\}: \psi_{\ell,1} < \ldots < \psi_{\ell,K^-_\ell}\right\}, \nonumber\\
\theta_{1}, \ldots, \theta_{L}  & \sim {\sf Beta}(N_{\theta} a_\theta,N_{\theta}(1-a_\theta))\ind\{(\max_{1\leq v \leq K^-_\ell}{\psi_{\ell, v}}, 1)\}, a_\psi  \sim {\sf Beta}(a_0, b_0), \text{~and~} a_\theta \sim {\sf Beta}(a_0', b_0'),\nonumber
\end{align}
for $\ell = 1, \ldots, L$, where $K^-_\ell\geq 1$ is the number of response probability parameters for latent classes $\bEta$ with $\Gamma_{\bEta, \ell}=0$ defined in (\ref{eq::design_mat_by_Q}) and the truncation of $\theta_{\ell}$ follows from the definition of RLCM (\ref{eq::rlcm_def}). With ($a_\theta$, $a_\psi$) unknown, the hierarchical priors on $\btheta$ and $\{\bpsi_v\}$ propagate into the posterior and have the effect of shrinking the parameters towards a population value by sharing information across dimensions; ($N_\theta$, $N_{\psi}$) can further be sampled in the posterior algorithm or fixed. When multi-parameter RLCMs specify particular parametric forms of the response probability for feature $\ell$ (e.g., in (\ref{eq::logit_cdm})), other sets of priors on the parameters may be readily incorporated into posterior sampling by modifying Step 4 in Supplementary Material C.1. Finally, we specify prior for  hyperparameter $\alpha_1$ in (\ref{eq::finite_ibp_step1}). One may specify a prior conjugate to $[H^* \mid \alpha_1]$ by $\alpha_1 \overset{d}{\sim} {\sf Gamma}(e_0,f_0)$ (shape and inverse scale parameterization with mean $e_0/f_0$ and variance $e_0/f_0^2$). Posterior sampling for non-conjugate prior for $\alpha_1$ can also be carried out by sampling over a dense grid upon bounded reparameterization (see Step 5 in Supplementary Material C.1).

Taken together, the likelihood and priors give the joint distribution of data $\mathbf{Y}=\{\bm{Y}_i\}$, the true and false positive rates $\btheta$ and $\bPsi$, ${Q}$ matrix, and latent state vectors ${H} = \{\bEta_i\}$ (see Supplementary Material A.8).

\section{Posterior Inference}
\label{sec::mcmc}

We design posterior sampling algorithms to address three questions, 1) how many scientific clusters $(\tilde{T})$ in the sample (data); 2) what are the latent state vectors $\{\tilde{\bEta}^*_j, j=1, \ldots, \tilde{T}\}$ in the sample; and 3) what are the subjects' latent states $\bEta_i$ and the scientific clusters $\tilde{\cC}$.  

Given $Q$, $\btheta$ and $\Psi$, RLCM as a mixture model has \textit{discrete} component-specific parameters $\bEta_i \in \cA$. This is to be contrasted with mixture models with a continuous base measure from which component parameters are drawn to differ from one another with probability one. Therefore, when sampled conditional on other parameters, the discrete component parameters $\{\bEta^*_j,j=1, \ldots, T\}$ may be duplicated. Because we are interested in estimating scientific clusters with distinct latent states, we \textit{post-process} the posterior samples by merging clusters in $\cC$ associated with identical $\bEta^*_{j}$ at each MCMC iteration. Given $M$, no more than $2^M$ distinct latent state vectors $\tilde{\bEta}_j^*$ results after merging. More generally, for inference based on mixture of finite mixture (MFM) models with \textit{discrete} component parameters, (\ref{eq::mixture1}) uses a prior over all non-negative integers to remove the otherwise hard constraint $K = \tilde{K} \leq 2^M$ (would be so if we force distinct latent states in the prior) and greatly simplify the design of posterior algorithms (see Remark {\ref{remark::pseudo_cluster}).

We use Markov chain Monte Carlo (MCMC) algorithm for posterior inference which by design simulate samples that approximate the joint posterior distribution of unknown parameters and latent variables: \((\bZ, H^*, Q, \btheta, \Psi, \alpha_1)\). See Supplementary Material C.1 for more details of the sampling algorithms and convergence checks. We discuss information from data that updates the clusters $\cC$.

\noindent \underline{\textit{Gibbs updates of the partitions}.} 
Given our focus on estimating clusters, we choose to directly sample $\mathcal{C}$ from its posterior without the need for considering component labels or empty components. A key step is to sample $\cC$ based on an urn process that begins with one cluster comprised of all subjects (or a warm start informed by crude initial clusters) and re-assigns each subject to an old or new cluster  \citep{miller2017mixture}. In sampling $\{Z_i\}$ one subject at a time, the full conditional distribution $[Z_i \mid \bm{Z}_{-i}, \mathbf{Y}, \btheta, \bPsi, Q, \bp]$ given cluster assignments for the rest $\bm{Z}_{-i}=\{Z_{i'},i'\neq i\}$, other model parameters and data is proportional to the product of the conditional prior $pr(Z_i \mid \bm{Z}_{-i}, \gamma)$ and the complete data likelihood integrated over latent states $[\mathbf{Y} \mid \bZ, \btheta, \Psi, Q, \bp]$ (equivalent to conditional upon partition $\mathcal{C}$ ignoring the labels). Because of exchangeability among subjects, we view subject $i$ as the last observation to be updated during a Gibbs step which assigns subject $i$ to an existing cluster $C \in \mathcal{C}_{-i}$ or a new cluster on its own with probabilities:
 \begin{align}
 \PP(Z_i = j \mid -) & \propto 
 \begin{cases}
 (|C|+\gamma)\cdot \frac{g(C\cup \{i\})}{g(C)}, &{\sf~if~} C\in \mathcal{C}_{-i}, j =1, \ldots,  |\mathcal{C}_{-i}|, {\sf ~or~} \\
 \gamma \frac{V_{N}(t+1)}{V_N(t)}\cdot  g(C), & {\sf~if~}C=\{i\}, j= |\mathcal{C}_{-i}|+1,
 \end{cases}\label{eq::gibbs_updates}
 \end{align}
 where $g(C) = g(C; \btheta, \Psi, Q, \bp) = \prod_{\ell=1}^L pr(\{Y_{i\ell}: i \in C\} \mid \btheta, \Psi, Q, \bp)$ is the marginal likelihood for data in cluster $C$ (see (S4) in Supplementary Material B for an illustration using model (\ref{eq::dino})). If adding subject $i$ to any existing cluster fits poorly with data $\mathbf{Y}_{C}$, i.e., knowing $\mathbf{Y}_{C}$ tells little about $\bY_i$, low marginal likelihood ratio $\frac{g(C\cup \{i\})}{g(C)g(\{i\})}$ will result for any $C\in \cC_{-i}$.  The Gibbs update will favor forming a cluster of its own $\{i\}$.
 
\noindent \underline{\textit{Posterior summaries}.} We summarize the posterior distribution of partitions $[\cC\mid \mathbf{Y}]$ by computing the empirical frequencies $\hat{ \pi}_{ii'}$ for every pair of subjects being clustered together, referred to as the posterior co-clustering probabilities $\pi_{ii'}=\PP(Z_i=Z_{i'} \mid \mathbf{Y})$, for subjects $i$, $i'=1, \ldots, N$.  We compute a simple least square (LS) clustering $\hat{\cC}^{(LS)}$ on the basis of the squared distance from the posterior co-clustering probabilities, $\arg \min _{b} \sum_{i,i'} \left\{\delta(Z_i^{(b)}, Z_{i'}^{(b)}) - \hat{\pi}_{ii'}\right\}^2$, where $\delta(a,a')=1$ if $a=a'$ and zero otherwise \citep{dahl2006model}. 

RLCM has the salient feature of subject-specific discrete latent states $\bEta_i$. However, the interpretation of $\bEta_i$ depends on $Q$ which is of scientific interest on its own in many applications. Based on the posterior samples obtained from a model with an unknown $Q$, we select the iteration(s) $b^*$ with the mininum loss, $ \min_{b^*} \|Q^{(b^*)\top} Q^{(b^*)}-\frac{1}{B}\sum_{b=1}^B Q^{(b)\top} Q^{(b)}\|_F$ where $\|A\|_F=\sqrt{\sum a^2_{ij}}$ is the matrix Frobenius norm.  $Q^\top Q$ is a $L$ by $L$ matrix invariant to relabeling of latent states. The $(\ell, \ell')$-th element of $Q^\top Q$ represents the number of activated states at feature $\ell$ when $\ell=\ell'$ and the number of co-activated states at feature pair ($\ell$, $\ell'$) when $\ell\neq \ell'$. Minimization of the least squares criterion therefore selects an iteration closest to the posterior means of all the co-activation counts. Turning to the inference of $\bEta_i$, although in the original MCMC chain the subset of the $H^{*(b)}$ and $\bZ^{(b)}$ samples drawn along with $Q^{(b^*)}$ usefully approximate $[\bEta_i = \bEta^*_{Z_i}, i=1, \ldots, N\mid Q=Q^{(b*)}, \mathbf{Y}]$, inferences of their functions enjoy reduced Monte Carlo errors through refitting a model with $Q=Q^{(b*)}$ that generate more posterior samples.  Section \ref{sec::analysis} further illustrates these use of the posterior summaries through detailed analyses of data from Example 1.

\section{Results}
\label{sec::results}

We illustrate the utility of RLCM on both simulated and real data. We focus on scenarios where $Q$ is unknown. First, we assess the performance of RLCM on estimating clusters under simulation scenarios corresponding to distinct levels of measurement errors, feature dimensions, sparsity levels of each machine, sample sizes, and population fractions of latent state patterns. Here the goal is to show that the proposed Bayesian RLCM performs clustering as well as or better than common alternative binary-data clustering methods. We first analyze a single randomly generated data set to highlight the differences among the methods. We then investigate the frequentist property of Bayesian RLCM in cluster estimation and compare it to other methods through repeated application of each method to replication data sets. Finally, data from Example 1 is analyzed, focusing on the posterior inferences of clusters, cluster-specific latent states and the estimated $Q$-matrix. 

\subsection{Simulated Examples to Study Model Performance}
\label{sec::simulation}

\noindent{\textit{Simulation 1: More accurate clustering through feature selection in scientifically structured classes.}} $N=50$ independent observations are generated from an $L=100$ dimension multivariate binary distribution with $M=3$ machines. Here we randomly generated an $M$ by $L$ matrix $Q$ where each row has on average $s=20\%$ non-zero elements. That is, $Q_{m\ell}\overset{i.i.d}{\sim}{\sf Bernouli}(0.2), \ell=1, \ldots, L$; In the rare event where a randomly generated $Q\notin \cQ$ (identifiability constraint (\ref{eq::identifying_constraint})), we randomly permute pairs of elements in $Q_{m\star}$ until $Q\in \cQ$. We draw latent states for each observation independently according to $\bEta_i \overset{d}{\sim}{\sf Categorical}\left(\bpi_0 = (1/6,1/6,1/6, 1/6,1/12,1/12,1/12, 1/12)\right)$ where 

\vspace{-1.2cm}
\[
\bpi_0 = \{\PP(\bEta_i = {\sf (0,0,0),(1,0,0),(0,1,0),(1,1,0),(0,0,1),(1,0,1),(0,1,1),(1,1,1)})\}.\]
\vspace{-1.2cm}

\noindent Here we focus on the two-parameter model ((\ref{eq::dino}), DINO) which will be applied to Example 1 in Section \ref{sec::analysis}. We assume the response probabilities shift between two levels $\theta_\ell = 0.8$ and $\psi_\ell=0.15$. The distinct subsets of features where shifts occur define eight classes $\tilde{K}=8=(2^M)$, which upon enumeration by observation gives an $N$ by $L$ design matrix $\Gamma$. The resulting data $\mathbf{Y}$, the design matrix $\Gamma$, as well as the clusters obtained using complete-linkage, Hamming distance hierarchical clustering ({\sf HC}), standard eight-class Bayesian latent class analysis ({\sf LCA}, e.g., \citet[][]{garrett2000latent}), subset clustering analysis \citep{hoff2005subset} and our Bayesian {\sf RLCM} with unknown number of clusters fitted with truncation level $M^\dagger=5$ can be seen in Figure \ref{fig::intro_motivating_example}. Specifically, for Bayesian {\sf LCA}, {\sf RLCM} and subset clustering \citep{hoff2005subset}, we plot the posterior co-clustering probability matrix $\{\hat{\pi}_{i,i'}\}$ for $N$ observations; For {\sf HC}, we indicate co-clustering by filled cells. The true clusters are separated (dashed grids) and ordered according to the truth. Filled blocks on the main diagonal indicate perfect recovery of the true clusters.  In this setting, {\sf HC} is sensitive to noise and tends to split a true cluster (blank cells within the main diagonal blocks) or group observations from different true clusters (blue cells in the off-diagonal blocks). Unlike the Bayesian {\sf LCA} and the subset clustering, the Bayesian {\sf RLCM} automatically selects and filter subsets of features that distinguish eight classes (through scientific structures in (\ref{eq::dino})) hence has superior clustering performance producing clusters that agrees quite well with the truth. This advantage of Bayesian {\sf RLCM} relative to alternatives is maintained under data replications (see Simulation 2).

\begin{figure}[h]
\captionsetup{width=0.95\linewidth}
\centering
\addtocounter{figure}{1} 
\includegraphics[width=0.9\linewidth]{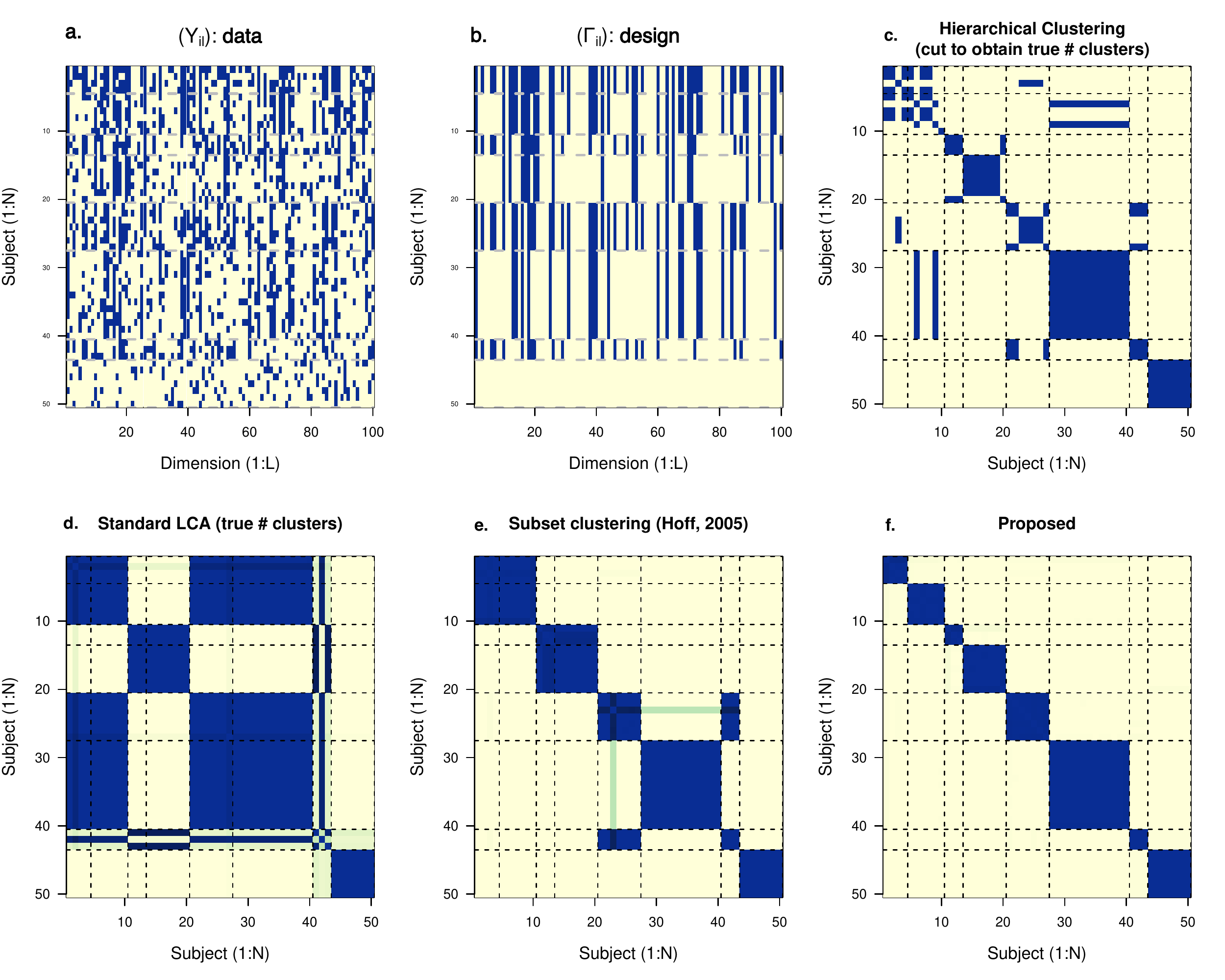}
\label{fig:truth}
\addtocounter{figure}{-1} 
\caption{In the 100-dimension multivariate binary data example, the eight classes differ with respect to subsets of measured features. Bayesian restricted latent class analysis accounts for measurement errors, selects the relevant feature subsets and filters the subsets by a low-dimensional model (\ref{eq::dino}) and therefore yields superior clustering results. }
\label{fig::intro_motivating_example}
\end{figure}

 Compared to traditional all-feature methods under large dimensions, through the inference of all-zero columns of $Q$ ($\{\ell: \Gamma_{\bEta, \ell}=0, \forall \bEta\in \cA \}$), Bayesian {\sf RLCM} removes irrelevant features hence reduces the impact of noise at less important features and in the current setting has better clustering performance (see Supplementary Material E for additional simulated examples on this point).

\noindent{\textit{Simulation 2: Assess clustering performance under various parameter settings.}} We simulated $R=60$ replication data sets for each of $1,920$ combinations of ({\sf \#features, sample~size, true~positive~rate, ~false~positive~rate, ~population~fractions, ~sparsity~level~of~the~rows~of~$Q$}): $(L,N,\theta_0,\psi_0,\bpi_0, s )\in \{50,100,200,400\} \otimes \{50,100,200\}\otimes \{0.8,0.9\}\otimes \{0.05,0.15\}\otimes \{\bpi_a = (\frac{1}{8}, \ldots, \frac{1}{8}), \bpi_b = (\frac{1}{6},\ldots, \frac{1}{6},\frac{1}{12},\ldots, \frac{1}{12})\}\otimes \{10\%,20\%\}$. The parameter values are designed to mimic what would be expected in Examples 1-3. 
We use adjusted Rand index  \citep[aRI,][]{hubert1985comparing} to assess the agreement between two clusterings, e.g,. the estimated and the true clusters. {\sf aRI} is defined by
\(
{\sf aRI}(\mathcal{C}, \mathcal{C}')  =  \frac{\sum_{r,c}{n_{rc}\choose 2}-\left[\sum_r {n_{r\cdot}\choose 2}\sum_c {n_{\cdot c}\choose 2}\right]/{N\choose 2}}{0.5\left[\sum_r {n_{r\cdot}\choose 2}+\sum_c {n_{\cdot c}\choose 2}\right]-\left[\sum_r {n_{r\cdot}\choose 2}\sum_c {n_{\cdot c}\choose 2}\right]/{N\choose 2}},
\)
where $n_{rc}$ represents the number of observations                                                                                                                                                                                                                                                                                                                                                                                                                                                                                                                                                                                                                                                                                                            placed in the $r$th cluster of the first partition $\mathcal{C}$ and in the $c$th cluster of the second partition $\mathcal{C}'$, $\sum_{r,c}{n_{rc}\choose 2} (\leq 0.5\left[\sum_r {n_{r\cdot}\choose 2}+\sum_c {n_{\cdot c}\choose 2}\right])$ is the number of observation pairs placed in the same cluster in both partitions and $\sum_r {n_{r\cdot}\choose 2}$ and $\sum_c {n_{\cdot c}\choose 2}$ calculates the number of pairs placed in the same cluster for the first and the same cluster for second partition, respectively. {\sf aRI} is bounded between $-1$ and $1$ and corrects for chance agreement. It equals one for identical clusterings and is on average zero for two random partitions; larger values indicate better agreements between the two clustering methods.

First we apply Bayesian {\sf RLCM} to each replication data set and focus on studying its performance in recovering the true clusters (boxes with solid lines in Figure \ref{fig:simulation_replication}). The clustering performance varies by the sparsity level  $(s)$ in each machine, level of measurement errors $(\theta_\ell, \psi_\ell)$, population fractions of latent classes $\{\pi_{\bEta}, \bEta\in \cA\}$ and sample sizes $(N)$. Given $s$, a larger $L$ means a larger number of relevant features per machine and leads to better cluster recovery. In Figure S2 of Supplementary Materials (Figure \ref{fig:simulation_replication} here shows its $8$ subplots), increasing $L$ from $50$ to $400$ (from the top to the bottom row), the mean {\sf aRI} (averaged over replications) increases, e.g., in the first column, from $0.7$ to $0.98$ at the sparsity level $s=10\%$, $0.88$ to $0.99$ under $s=20\%$. More generally, clustering performance improves by increasing the sparsity level in each machine from $s=10\%$ to $20\%$ (compare the 1st and 3rd, 2nd and 4th {\sf RLCM} boxplots with solid lines in each panel of Figure \ref{fig:simulation_replication}). In the context of Example 1, given a fixed number of protein landmarks $L$, patients will be more accurately clustered if each machine comprises more component proteins. This observation is also consistent with simulation studies conducted in the special case of $Q=I_L$ \citep[][Table 1]{hoff2005subset}. 

We obtain more accurate cluster estimates under larger discrepancies between $\theta_\ell$ and $\psi_\ell$. For $\theta_0$ fixed at $0.8$ or $0.9$, the mean {\sf aRI} averaged over replications is higher under $\psi_0=0.05$ than $\psi_0=0.15$ over all combinations of the rest of parameters. Under the non-uniform population fraction $\bpi_0=\bpi_b$, the clustering performance by Bayesian {\sf RLCM} is similar or slightly worse than under a uniformly distributed population ($\bpi_a$). Finally, we observe mixed relative performances at distinct sample sizes as a result of two competing factors: more precise estimation of measurement error parameters under large sample sizes that improve clustering and a larger space of clusterings under a larger $N$.

Figure \ref{fig:simulation_replication} also shows better clustering performance of Bayesian {\sf RLCM} (boxes with solid lines) relative to the three common alternatives (boxes with dotted lines). The Bayesian {\sf RLCM} on average most accurately recovers the clusters compared to other methods. Bayesian {\sf RLCM} produces the highest {\sf aRI}s compared to others which are in many settings perfect (close to one). For example, the ratio of the mean {\sf aRI}s (averaged over replications) for Bayesian {\sf RLCM} relative to subset clustering is $2.06$, $2.04$, $1.88$, $1.71$ for the sample-size-to-dimension ratios $N/P=1, 0.5,0.25,0.125$, respectively (the leftmost group of four boxplots in Column 1, Figure S2 of Supplementary Materials $\psi_0=0.05$, $s=10\%$, $\bpi_0=\bpi_a$); The relative advantage of Bayesian {\sf RLCM} and {\sf HC} narrows under a higher false positive rate ($\psi_0=0.15$) as shown by the smaller {\sf aRI} ratios $1.23$, $1.62$, $1.49$, $1.16$  (the leftmost group of four boxplots in Column Two, Figure S2).

We remark on the performance of other three methods. Over all parameter settings investigated here, the traditional {\sf LCA} performed the worst in the recovery of true clusters ({\sf aRI}  $< 0.68$). The likelihood function of subset clustering is a special case of {\sf RLCM} that assumes a non-parsimonious $Q=I_L$ and therefore loses power for detecting clusters compared to {\sf RLCM} that estimates a structured $Q$ with multiple non-zero elements in its rows. {\sf HC} is fast and recovers the true clusters reasonably well (ranked second or first among the four methods  more than two thirds of the parameter settings here; See Figure S3 in Supplementary Materials). The performance of {\sf HC} is particularly good under a low level of measurement errors ($\psi_0=0.05$) and a large number of relevant features per machine and sometimes performs much better than traditional {\sf LCA} and subset clustering (e.g., $L=200$, $N=50$, $\theta_\ell=0.8$, $\psi_\ell=0.05$ in Figure S2, Supplementary Materials). The {\sf HC} studied here requires a pre-specified number of clusters to cut the dendrogram at an appropriate level and produces clusters that require separate methods for uncertainty assessment \citep[e.g.,][]{suzuki2006pvclust}. The proposed Bayesian {\sf RLCM}, in contrast, enjoys superior clustering performance and provides direct internal assessment of the uncertainty of clusters and measurement error parameters through the posterior distribution. 

\begin{figure}[h]
\captionsetup{width=\linewidth}
\centering

\includegraphics[width=0.75\linewidth]{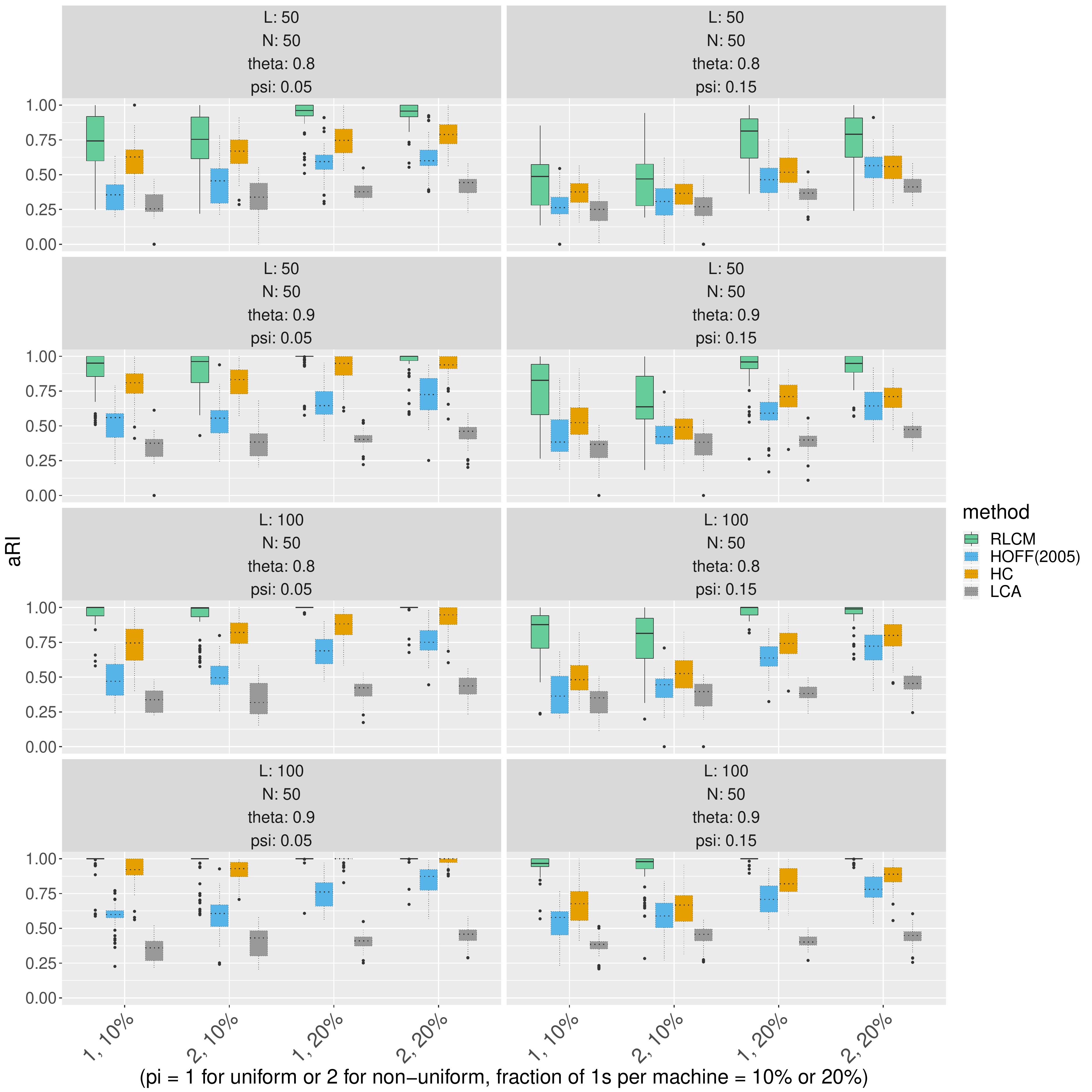}
\label{fig:simulation_replication}

\caption{Based on $R=60$ replications for each parameter setting, Bayesian {\sf RLCM} (boxplots with solid lines) most accurately recovers the true clusters compared to subset clustering (Hoff, 2005) hierarchical clustering ({\sf HC}) and traditional Bayesian latent class analysis ({\sf LCA}) (from the left to the right in each group of four boxplots). See Figure S2 in Suppmentary Materials for an expanded version over more parameter settings.}
\label{fig:simulation_replication}
\end{figure}

\subsection{Analysis of GEA Data}
\label{sec::analysis}

%
%
%
%
%
%

\subsubsection{GEA Data, Preprocessing and Informative Priors}
Example 1 is about estimating autoimmune disease patient clusters via reconstructing components of protein complexes. Autoantibodies are the immune system's response to specific cellular protein complexes or ``machines". We seek to identify components of the machines and to quantify the variations in their occurrence among individuals. The binary responses $\bY_i$ indicate the observed presence of autoantibodies at equi-spaced molecular weight landmarks as produced via a preprocessing method \citep{wu2017gelpre} implemented using publicly available software \verb"R" package ``\verb"spotgear"" (\href{https://github.com/zhenkewu/spotgear}{https://github.com/zhenkewu/spotgear}). We ran $4$ GEA gels, each loaded with IPs performed using sera from $19$ different patients, and one reference lane. All sera were from scleroderma patients with cancer, and were all negative for the three most common autoantibodies found in scleroderma (anti-RNA polymerase III, anti-topoisomerase I, and anti-centromere). The IPs were loaded in random order on each gel; the reference sample is comprised of known molecules of defined sizes (molecular weights) and was always loaded in the first lane. The left panel in Figure \ref{fig:example1_ssc_results} shows for each sample lane (labeled in the left margin; excluding the reference lanes) the binary responses indicating the observed presence or absence of autoantibodies at $L=50$ landmarks. 

Patients differ in their antibody protein presence or absence patterns at the protein landmarks. Eleven out of $L=50$ aligned landmarks are absent among the patients tested. The rest of the landmarks are observed with prevalences between $1.3\%$ and $94.7\%$. We apply two-parameter RLCM (\ref{eq::dino}) with unknown $M(<L/2=50)$ and $Q$, $\btheta$, $\bpsi$. The GEA technologies are known to be highly specific and sensitive for nearly all proteins studied in this assay so we specify the priors for the true and false positive rates by ${\sf Beta}(a_{\theta\ell},b_{\theta\ell})$ and ${\sf Beta}(a_{\psi\ell},b_{\psi\ell})$, $\ell = 1, \ldots, L$ respectively.  We set $a_{\theta\ell}=9$, $b_{\theta\ell}=1$, $a_{\psi\ell}=1$, $b_{\psi\ell}=99$ and conducted sensitivity analyses varying these hyperparameter values. Because proteins of distinct weights may have systematically different measurement errors, we choose not to share measurement error rates across dimension in this analysis. In our analysis, we sampled many $Q$ across iterations of MCMC. Because the interpretation of $\bEta_i$ depends on the row patterns in $Q$, we condition on the least square clustering ($\hat{\cC}^{(LS)}$) and refit the model to obtain the least square $Q$ (Section \ref{sec::mcmc}). The prior of $H^*$ (Section \ref{sec::prior_H}) prevents overfitting by encouraging a small number of \textit{active} latent states ($\{m: \sum_{i}\eta_{im}\neq 0\}$) for small $\alpha_1$ which in this analysis we draw  its posterior samples for inference.

{In this application, the scientists had previously identified and independently verified through additional protein chemistry the importance of a small subset of protein bands in determining clusters. They proposed that these proteins should be grouped together. We therefore fitted the Bayesian RLCM without further splitting these partial clusters $\cC^{(0)}$ so that the number of scientific clusters visited by the MCMC chain has an upper bound $\tilde{T}^{(b)}\leq |\cC^{(0)}|+N-\sum_{j=1}^{|\cC^{(0)}|}C^{(0)}_j$, where $C^{(0)}_j$ counts the number of observations in the initial cluster $j$. We fitted models and compared the results under multiple ``working" truncation levels $M^\dagger=8,9,\ldots, 15$ and obtained identical clustering results.}

\subsubsection{GEA Results}
Figure \ref{fig:example1_ssc_results} shows: the observations grouped by the RLCM-estimated clusters (not merged) $\hat{\cC}^{(LS)}$ (left), the estimated $Q$-matrix $\hat{Q}(\hat{\cC}^{(LS)})$ (right), and the marginal posterior probabilities of the machines $\PP(\eta_{im}=1 \mid \hat{\cC}^{(LS)}, \hat{Q}(\hat{\cC}^{(LS)}), \mathbf{Y})$ (middle). 

The matrix $Q$ is estimated from the observed marginal associations (positive or negative) among the protein landmarks.  Landmark protein pairs observed with \textit{positive} association tend to be placed in  the same estimated machine. For example, Landmarks 4, 7 and 8 appear together in Machine 5. Subjects either have all three landmarks or none at all, which induces strong positive pairwise associations among these landmarks. Indeed, the estimated log  odds ratio (LOR) is $3.13$ (standard error $1.16$) for Landmark 4 versus 7, $2.21$ (s.e., $0.98$) for Landmark 4 versus 8, and $2.92$ (s.e. $1.2$) for Landmark 7 versus 8. 

The observed \textit{negative} marginal associations between two landmarks suggest existence of machines with discordant landmarks. For example, Landmarks 10 and 27 are rarely estimated to be present or absent together in a subject as a result of 1) estimated machines with discordant landmarks and 2) subject-specific machine assignments. First, the model estimated that Landmark 10 (in {\sf Machine Set A}: 1, 3 and 4) belongs to machines not having Landmark 27 (it is in {\sf Machine Set B}: 2). Second, with high posterior probabilities, most observations  have machines from one of, not both Set A and B hence creating  discordance (high posterior probability $\PP(\Gamma_{i,10}\neq \Gamma_{i,27}\mid \mathbf{Y})$). In the presence of observation errors, strong negative marginal association results (observed LOR for Landmark 10 versus 27: $-1.98$, s.e. $0.8$).

\begin{figure}[h]
\captionsetup{width=\linewidth}
\centering

\includegraphics[width=\linewidth]{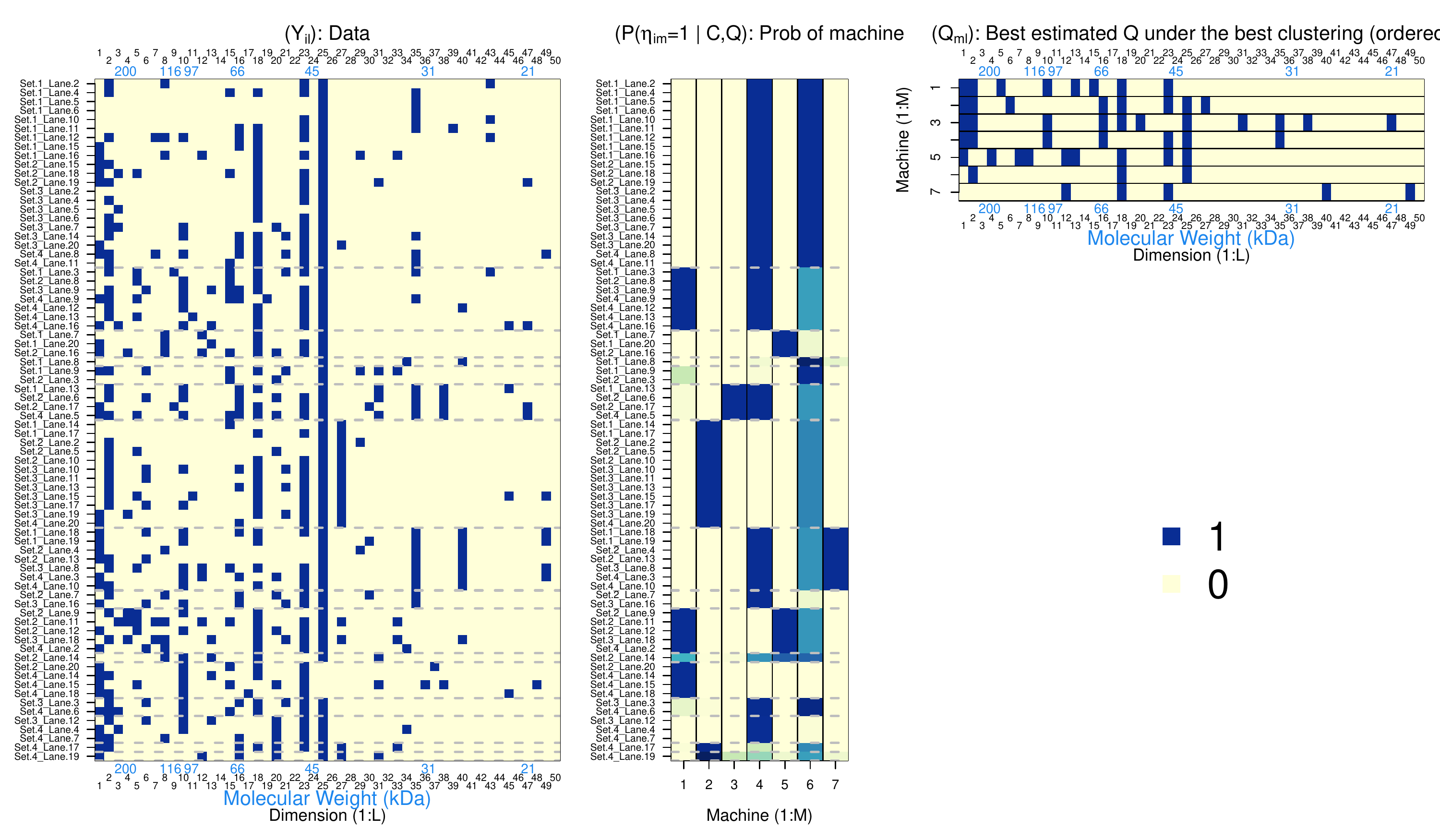}
\label{fig:example1_ssc_decomp}

\caption{Results for GEA data in Example 1.
 \textit{Left}) Aligned data matrix for band presence or absence; row for $76$ serum lanes, reordered into optimal estimated clusters (not merged) $\hat{\cC}^{(LS)}$ separated by gray horizontal lines ``-----”; columns for $L=50$ protein landmarks. A blue vertical line ``{\color{blue} $\bm{\vert}$}" indicates a band;
\textit{Middle}) lane-machine matrix for the probability of a lane (serum sample) having a particular machine. The blue cells correspond to high probability of having a machine in that column. Smaller probabilities are shown in lighter blue;.
\textit{Right}) The estimated machine profiles. Here seven estimated machines are shown, each with component proteins shown by a blue bar ``{\color{blue} $\vert$}". }
\label{fig:example1_ssc_results}
\end{figure}

Our algorithm also directly infers the number of scientific clusters in the data given an initial partial clustering $\cC^{(0)}$. The marginal posterior of the number of scientific clusters $\tilde{T}$ can be approximated by empirical samples of $\{\tilde{T}^{(b)}\}$ which result in a posterior median of $12$ ($95\%$ credible interval: $(8,16)$; Figure S4 in Supplementary Materials). The advantage of Bayesian RLCM is the posterior inference about both the clusters and the distinct latent state variables $\bEta_i$ interpreted based on the inferred $Q$ matrix. The middle panel of Figure \ref{fig:example1_ssc_results} shows that clusters differ in their marginal posterior probabilities of having each of the estimated machines. Among $76$ subjects analyzed,  $23$ of them have  greater than $95\%$  marginal posterior probabilities of having both Machine 4 and 6. A group of seven observations are enriched with Machine 4 and 7 which as expected from the raw band patterns have distinctive combination of Landmarks 35, 40 and 49 (33, 27 and 18 kDa bands, respectively). Such inference about $\bEta_i$ is not available to us based on hierarchical clustering or traditional latent class models. 

{ We also fitted a Bayesian RLCM without the partial clusters $\cC^{(0)}$ identified in prior work by the scientists. We estimated lower true positive rates so that it is more likely to observe negative protein landmarks within clusters partially identified by having a machine with a protein at that landmark. This makes the findings more difficult to interpret. As discussed in the simulation studies, clustering performance of Bayesian RLCM is poorer under lower sparsity levels $s=10\%$. As our scientific team recruits and tests more serum samples from their scleroderma patient cohort,  samples with novel antibodies will improve inference about the measurement error parameters. This highlights the importance of using available  prior knowledge about the measurement technologies in inferring latent states in finite samples \citep[e.g.,][]{wu2016partially}. Figure S5 in Supplementary Materials compares for each landmark the prior and posterior distributions of the true and false positive rates. The discrepancies observed at many landmarks suggest the learning of measurement error parameters from the data. Other landmarks have similar prior and posterior distributions as a result of nearly flat likelihood function or absence of protein at that landmark so learning based only on likelihood is  impossible. }

We performed posterior predictive checking to assess model fit \citep{gelman1996posterior}. At each MCMC iteration, given the posterior sample of model parameters (without conditioning on the best clustering $\hat{C}^{(LS)}$ or the best $\hat{Q}$), we simulated a data set of the same size as the original set. For each replicated data set, we compute the marginal means and marginal pairwise log odds ratios ($0.5$ adjustment for zero counts). Across all replications, we compute the $95\%$ posterior predictive confidence intervals (PPCI) defined by the $2.5\%$ and $97.5\%$ quantiles of the PPD. All the observed marginal means are covered by their respective PPCIs; The $95\%$ PPCIs cover all but $24$ of ${L \choose 2}=1,225$ landmark pairs of observed pairwise log odds ratios (see Figure S6 and S7 in Supplementary Materials). The proposed model adequately fits the GEA data. 

There are potential improvements in our analysis. The posterior predictive probabilities (PPP) of observing a more extreme log odds ratio in future data $\PP({\sf LOR_{1,2}}(\mathbf{Y}^{\sf rep}) < {\sf LOR_{1,2}}(\mathbf{Y}) \mid \mathbf{Y})$ are between $0.004$ and $0.024$. Most of these  misfits of marginal log odds ratio occurred for landmark pairs with an observed marginal two-way table with small cell counts. Because the Bayesian RLCM treats the zeros as random, if these zero cells correspond to impossible combinations of proteins, or structural zeros, it may overestimate the probability for these cells; See \citet{manrique2014bayesian} for a truncated extension of traditional latent class models that can be adapted to address the structural zero issue. On the other hand, the neighboring Landmarks 1 and 2 have an observed log odds ratio of $-1.17$ (s.e. $0.48$) with PPP $0.011$. The two landmarks compete for being aligned with an observed band during pre-processing \citep{wu2017gelpre} hence creating negative dependence even within a latent class. Deviation from local independence can be further accounted for by explicitly modeling local dependence structure, discussed elsewhere, e.g., by nesting subclasses within each class \citep[e.g.,][]{wu2017nplcm}.

\section{Discussion}
\label{sec::discussion}

Modern scientific technologies give rise to measurements of varying precision and accuracy that are better targeted at the underlying state variables than ever before. In this paper we have discussed Bayesian restricted latent class model for analyzing multivariate binary data in the presence of between-class differential errors. The focus has been on the clustering of observations with unknown number of clusters,  uncertainty assessment of the clustering and the prediction of individual latent states. The proposed method is motivated by clustering autoimmune disease patients based on their antibody presence or absence in sera where it is scientifically meaningful to \textit{restrict} the values of response probabilities among latent classes. We have compared the proposed method with variants of latent class models through their specifications in Table S1 in Supplementary Materials and illustrated its advantage through simulations relative to three commonly used binary-data clustering. The Bayesian RLCM performs what we have called scientifically-structured clustering. It automatically selects subset of features for each latent class and filters them through a low dimensional model to improve our ability to accurately estimate clusters. Though the present paper focused on demonstrating the method through an example in medicine, the developed method and algorithms apply to many problems including Example 2 and 3 (Section \ref{sec::intro}).  

RLCMs decompose the variation among multivariate binary responses into structure that reflects prior scientific knowledge and stochastic variation without a known explanation. In Example 1, it is certainly likely that there is some variability related to the vagaries of the measurement assay. However, it is also highly likely that there are systematic biological and biochemical processes not included in the structural part because they are unknown to us today.  RLCM analyses can be a useful tool in the effort to uncover the unknown structure. One approach would be to show that the latent classes are diagnostic of specific diseases. Another is that we might uncover a novel mechanism by defining distinct patterns of the same autoantigen machine in patients with the same disease or potentially in patients with different diseases that target the same machines.

This paper has focused on developing and applying RLCMs and algorithms to identify clusters and estimate subject-specific latent states. However, applied to public health research (e.g., pneumonia etiology research in Example 3), RLCM analyses more often focus on population quantities such as $\Pi=\{\pi_{\bEta}, \sum_{j}\pi_{\bEta} = 1, \bpi_{\bEta}\geq 0, \bEta \in \{0,1\}^M\}$ an $M$-way contingency table characterizing the population frequencies of the latent state vector $\bEta_i$.  Further research into flexible and parsimonious parameterization of $\Pi$ and its regression formulation in RLCMs are warranted. For example, quadratic exponential family \citep{zhao1990correlated} with negative second-order natural parameters assigns higher probabilities for $\bEta$ comprised of  few ones or use another level of latent Gaussian variables to induce flexible dependence among $\bEta_i$ \citep[e.g.,][]{xu2009probit}.

We are currently studying a few potentially useful model extensions. First, nested partially LCMs \citep{wu2017nplcm} incorporate local dependence and multiple sensitivity parameters $(K^+>1)$ that would improve the utility of Bayesian RLCMs as well. Second, because the algorithm involves iterating over subjects to find clusters in (\ref{eq::gibbs_updates}), the computational time increases with the number of subjects $N$. Divide-Cluster-Combine schemes that estimate clusters in subsamples which are then combined may improve the computational speed at the expense of the approximation introduced by the multi-stage clustering \citep{ni2018parallel}. Finally, in applications where the clustering of multivariate binary data comprises an important component of a hierarchical Bayesian model with multiple components, the posterior uncertainty in clustering propagates into other parts of the model and can be integrated into posterior inference of other model parameters \citep[e.g.,][]{jacob2017better}.

\vspace{-.5cm}
\section*{Software Availability}
\label{sec::software}

\vspace{-.25cm}
All model estimations are performed by an \verb"R" package ``\verb"rewind"", which is freely available at \href{https://github.com/zhenkewu/rewind}{https://github.com/zhenkewu/rewind}. 

\vspace{-.5cm}
\section*{Supplementary Materials}

\vspace{-.25cm}
The supplementary materials contain referenced figures, a table, remarks, and further technical details, e.g., on identifiability and sampling algorithms, as well as additional simulations and extended data analysis results.

\vspace{-.5cm}
\section*{Acknowledgment}
                                                                                                                                                                                                                                                                                                                                                                                                                                                                                                                                                                                                                                                                                                               
                                                                                                                                                                                                                                                                                                                                                                                                                                                                                                                                                                                                                                                                                                               \vspace{-.25cm} 
                                                                                                                                                                                                                                                                                                                                                                                                                                                                                                                                                                                                                                                                                                         The research is supported in part by a gift from the Jerome L. Greene Foundation and by the Patient-Centered Outcomes Research Institute (PCORI) Award (ME-1408-20318), National Institutes of Health (NIH) grants R01 AR073208, P30 AR070254 and P30 CA-046592 (ZW, Cancer Center Support Grant (CCSG) Development Funds from University of Michigan Comprehensive Cancer Center (UMCCC)). We also thank Gongjun Xu, Peter Hoff and Jian Kang for their insightful comments.
                                                                                                                                                                                                                                                                                                                                                                                                                                                                                                                                                                                                                                                                                                             
                                                                                                                                                                                                                                                                                                                                                                                                                                                                                                                                                                                                                                                                                                                \bibliography{gel_bmf}
\bibliographystyle{apalike}

\newpage
\appendix
\begin{center}
		{\large \bf Supplementary Materials for ``A Bayesian Approach to Restricted Latent Class Models for Scientifically-Structured Clustering of Multivariate Binary Outcomes"}
		\end{center}

\renewcommand{\thefigure}{S\arabic{figure}}
\renewcommand{\theequation}{S\arabic{equation}}
\renewcommand{\thetable}{S\arabic{table}}
\renewcommand{\thetheorem}{S\arabic{theorem}}

\setcounter{figure}{1}
\setcounter{equation}{1}
\setcounter{table}{1}
\addtocounter{theorem}{-1}

The supplementary materials contain referenced remarks, figures and a table in Main Paper, and further technical details, e.g., on identifiability and sampling algorithms, as well as additional simulations and extended data analysis results. In particular, Section A contains remarks, Section B illustrates the calculation of marginal likelihood central to the posterior sampling of clusters ((15) in Main Paper), Section C details the posterior algorithms for pre-specified $M$ (Section C.1) and infinite $M$ (Section C.2), respectively. Section D briefly summarizes useful theoretical identifiability conditions for RLCMs based on \citet{gu2018}. Section E illustrates through simulations the benefit of removing irrelevant features. Finally, Section F collects a table for variants of LCMs as well as figures for model results on the data analysis in Main Paper.
                                                                                                                                                                                                                                                                                                                                                                                                                                                                                                                                                                                                                                                                                                               
\section{Remarks }
\subsection{On Extending Prior  of $H$ to $M=\infty$}
In Main Paper, we have focused on models with a finite number of latent states with $M=M^\dagger$ typically set to a number that is large enough for the particular applications. In the MCMC sampling (Supplementary Material \ref{sec::split_merge}), not all of the ``working" $M^\dagger$ states will be used by the observations. The active number of states is usually strictly smaller than $M^\dagger$ based on simulations. We extend to infinite $M$ to obtain a prior for $H^*$ under infinite dimension of latent state vectors ($\bEta_i$). We take $M$ in (14) in Main Paper to infinity and obtain infinite-column prior for $H$ (through a prior on $H^*$ in Section 2.6.2 in Main Paper); This construction defines the infinite Indian Buffet process \citep{ghahramani2006infinite}. Supplementary Material  \ref{sec::unknown_M} provides posterior sampling algorithms for dealing with an infinite number of latent states by a novel slice sampler without the need of truncation \citep{teh2007stick}. 

\subsection{RLCM Connection to \citet{hoff2005subset}}
Setting $Q=I_{L\times L}$ and $\bEta_i \in \cA = \{0,1\}^L$ (i.e., $M=L$) gives ``mixture of Bernouli products" with each latent class (defined by $\bEta_i$) having \textit{relevant} features at possibly overlapping subsets of features $\cL_{\bEta} = \{\ell: \Gamma_{\bEta,\ell}=1\}$, $\bEta\in \cA$ \citep{hoff2005subset}. \citet{hoff2005subset} assumes the positive response probability $\lambda_{i\ell} =  \left\{\theta_{\ell,v}\right\}^{\Gamma_{i\ell}}(\psi_{\ell})^{1-\Gamma_{i\ell}}$, where $\Gamma_{i\ell}= \eta_{i\ell}$ given $Q=I_{L\times L}$ and the multiple true positive rates $\{\theta_{\ell,v}\}$ are greater than a single false positive rate $\psi_\ell$, for $\ell=1, \ldots, L$. This model can be written into a RLCM form with $K^+=1$ and $K^-\geq 1$ by reparametrization: $\Gamma^*_{i\ell}=1-\Gamma_{i\ell}$, $\psi^*_{\ell,v} = 1-\theta_{\ell,v}$ and $\theta^*_{\ell}=1-\psi_{\ell}$ and relabeling of the outcomes $Y^*_{i\ell}=1-Y_{i\ell}$. Indeed, the positive response probability under relabeling and reparameterization is $\lambda^*_{i\ell} = \PP(Y^*_{i\ell}=1 \mid - ) =  1-\PP(Y_{i\ell}=1 \mid - ) = 1-\lambda_{i\ell} = \left\{\psi^*_{\ell,v}\right\}^{1-\Gamma^*_{i\ell}}(\theta^*_{\ell})^{\Gamma^*_{i\ell}}$. 

\subsection{Additional Identifiability Considerations for Designing Posterior Algorithms}
We now turn to inferring subject-specific latent state vectors $ H = \{\bEta_i\}$ based on \textit{complete-data} likelihood $[\{\bY_i\} \mid H, \Lambda,Q]$. Even given $Q$, conditions for identifying $H$ exist but may fall short of ensuring consistent estimation of $H$ because the number of unknowns in $H$ diverges as the sample size increases. For example, it requires extra conditions that the number of measurements $L$ increases with the sample size \citep[e.g.,][]{chiu2009cluster}. In finite samples and dimensions, we address this issue in a Bayesian framework by \textit{in a priori} encouraging $H$ to be of low complexity, i.e., few clusters of distinct and sparse latent state vectors $\{\bEta_i\}$, which combined with data likelihood will by design tend to concentrate the posterior at such low-complexity $H$. 

In addition, when the latent space $\cA \subsetneqq \{0,1\}^M$, general identifiability theory for $Q$ depends on the identifiability of $\Gamma$, the structure of which then determines the set of $Q$s that are identifiable from the observed data distribution. Some RLCMs motivate our posterior algorithm design. For example, in two-parameter RLCMs, if two latent states are either always present or absent at the same time (``partners"), it is impossible for the likelihood alone to distinguish it from a model that combines the two latent states. In our posterior algorithm, we therefore merge such ``partner" latent states if present at some iterations and the corresponding rows in $Q$ (Step 3, Supplementary Material \ref{sec::split_merge}).  As another example, two latent states can form a hierarchical structure, that is, one latent state cannot be present unless the other is. Suppose the second latent state require the first latent state, then $Q_{2\ast}$ values at $\{\ell: Q_{1\ell}=1\}$ can be zero or one without altering the model likelihood. The sparsity priors on $H$ and the rows of $Q$ constraining $\sum_{\ell} Q_{m\ell}$ therefore concentrate the posterior distributions of $H$ and $Q$ towards low-dimensional latent states and a smaller number of rows in $Q$ (Section 2.6.2 in Main Paper). 


\subsection{{Prior information about $\Lambda$.}}
In applications where prior information about a subset of response probabilities $\Lambda$ is available, it is essential to integrate the informative priors into model estimation if strict or generic identifiabilities do not hold  \citep[e.g.,][]{gustafson2009limits, wu2016partially}. The sufficient conditions (C1) and (C2) in Main Paper ensure identifiability of $Q$ with completely unknown $(\Lambda, \bpi_{\tilde{K}})$.  Otherwise, absent likelihood-based identifiability of $Q$ and other parameters, prior information about $\Lambda$ alleviates the non-identifiability issue by concentrating the posterior at parameter values that better explain the observed data in light of the informative priors. In general non-identified models, the uncertainty in the prior will propagate into the posterior and will not vanish even as the sample size approaches infinity \citep[e.g.,][]{Kadane1974}.

\subsection{Prior for Partition $\cC$}

The prior distribution $p(\mathcal{C}\mid \gamma, p_K(\cdot))$ is an exchangeable partition probability function \citep[EPPF,][]{pitman1995exchangeable}, because it only symmetrically depends on the sizes of each block of the partition $\{|\cC_j|: \cC_j \in \cC\}$. \citet[][Theorem 4.1]{miller2017mixture} also derives an \textit{urn process} for generating partitions $\mathcal{C}_1, \mathcal{C}_2, \ldots, $ such that the probability mass function for $\cC_N$ is given by \(
p(\mathcal{C}\mid \gamma, p_K(\cdot)) = V_N(T)\prod_{C\in \mathcal{C}} \gamma^{(|C|)}, 
\); we will use this urn process for Gibbs updates of $\{Z_i\}$ one subject at a time in (17) in Main Paper. Note that the mapping from $\bm{Z}$ to $\cC$ is many-to-one with each $\cC$ corresponding to ${ K \choose T}T!$ distinct $\bm{Z}$ that differ by relabeling. Starting from a prior for partition $\cC$ then followed by drawing component-specific parameters from their prior distributions is particularly fruitful in product partition models \citep[e.g.,][]{hartigan1990partition}.

\subsection{On Merging Clusters with Identical Discrete Latent States}
At each MCMC iteration, two observations falling in distinct clusters ($Z_{i}\neq Z_{i'}$) might have identical latent states, i.e., $\bEta^*_{Z_i} = \bEta^*_{Z_i'}$  where the equality holds elementwise. At each iteration, we use unique multivariate binary vectors among all subjects $H = \{\bEta_i = \bEta^*_{Z_i}, i=1,\ldots, N\}$ to define ``scientific clusters" $\tilde{\mathcal{C}}$ through merging clusters associated with identical latent states. That is, 
\[\tilde{\mathcal{C}}= \left\{\{i: \bEta_i = \tilde{\bEta}^*_j\}, j = 1, \ldots, \tilde{T} \right\}\] 
where $\{\tilde{\bEta}^*_j, j=1,\ldots,\tilde{T}\}$ collects $\tilde{T}(\leq T)$ unique patterns among $\{\bEta^*_j, j=1, \ldots, T\}$. Let $\mathcal{M}: \{\bEta^*_{Z_i}, i=1, \ldots, N\}\mapsto \tilde{\mathcal{C}}$ represent this merge operation, i.e., $\tilde{\cC} = \cM(\{\bEta^*_j\},\{Z_i\})$.

As detailed in Section 3 in Main Paper, we first build on Gibbs updates (15) and split-merge updates \citep[e.g.,][]{jain2004split} to efficiently sample $\{Z_i\}$ from its posterior distribution. Given $\{Z_i\}$, we then update $H^*=\{\bEta^*_j\}$ and merge clusters $\cC$ to obtain $\tilde{\cC}$ via the mapping $\cM$. Define partial ordering $``\preceq"$ over partitions $\cC_1 \preceq \cC_2$ if  for any $C_1 \in \cC_1$, one can find a $C_2 \in \cC_2$ satisfying $C_1\subseteq C_2$. We have  $\cC\preceq \tilde{\cC}$, i.e., $\tilde{\mathcal{C}}$ is coarser than $\mathcal{C}$. Our procedure for obtaining clusters $\tilde{\cC}$ differs from mixture models where distinct $Z_i$ values with probability one correspond to distinct component parameters sampled from a continuous  base measure \citep[e.g.,][Proof of Theorem 4.2]{miller2017mixture}. $\tilde{\mathcal{C}} = \mathcal{C}$ is implicitly assumed in \citet{hoff2005subset} under a Dirichlet process mixture model. 

We specify priors on $K$ that represents the distinct values that $\{Z_i\}$ can take and a prior on $H^* = \{\bEta^*_j,j=1, \ldots, T\}$, which together induce a prior for $\tilde{\mathcal{C}}$ via 
\begin{align}
p(\tilde{\mathcal{C}}\mid \alpha_1, \gamma) & = \sum_{\cC: \cC\preceq \tilde{\cC} }p(\tilde{\cC}\mid \cC, \alpha)\cdot p(\cC \mid \gamma) \\
& =\sum_{\cC: \cC\preceq \tilde{\cC}} {2^M \choose \tilde{T}} (\tilde{T})! \left\{\int p(H^*\mid \cS, \bp)p(\bp\mid \alpha_1) \mathrm{d}\bp\right\} \cdot p(\cS \mid \gamma)\cdot T!,
\end{align}
where $\cS = \{S_1, \ldots, S_T\}$ is a ordered partition of $N$ subjects, obtained by randomly ordering parts or blocks of $\cC$ uniformly over $T!$ possible choices and $p(\cS \mid \gamma) \cdot T! = p(\cC \mid \gamma)$.

The prior for the number of components $K$ serves to regularize the number of clusters $T = |\mathcal{C}|$ among observed subjects (see \citet[][Equation 3.6]{miller2017mixture}). Because $\tilde{\cC}$ is coarser than $\mathcal{C}$,  a exponentially decaying prior on $K$ then encourages a small number of scientific clusters $\tilde{\cC}$ among $N$ subjects which results in using fewer component specific parameters to fit finite samples and improves estimation of unknown $H^*$ and $Q$.

\subsection{On Prior for $Q$}
In applications where $Q$ is not fully identifiable or encouraged to be different among its rows, we specify sparsity priors for each column of $Q$ to encourage proteins to be specific to a small number of machines. That is, $\PP(Q_{m\ell} \mid \{Q_{m',\ell},m'\neq m\},  \zeta) = 1/\left\{1+\exp\left\{- \zeta \sum_{1\leq m' < m'' \leq M^*}Q_{m'\ell}Q_{m''\ell}\right\}\right\}$, where $ \zeta$ is the canonical parameter characterizing the strength and direction of interactions among $m$. We either fix $\zeta$ to be a negative number, or specify a hyperprior for $\zeta$; In this paper, we fix $\zeta=0$.

\subsection{Joint Distribution}

The joint distribution of data $\mathbf{Y}=\{\bm{Y}_i\}$, true and false positive rates $\btheta$ and $\bPsi$, ${Q}$ matrix, and latent state vectors ${H} = \{\bEta_i\}$, denoted by ${pr(\bY,{H = H(H^*,\bZ)},{Q}, \btheta, \bPsi)}$, is
\begin{align}
 &\left\{\prod_{i=1}^N \prod_{\ell=1}^L \left[\Gamma_{\bEta_i, \ell} \theta_{\ell}^{Y_{i\ell}}(1-\theta_{\ell})^{1-Y_{i\ell}}+(1-\Gamma_{\bEta_i, \ell})\psi_{\ell, v_i}^{Y_{i\ell}}(1-\psi_{\ell, v_i})^{1-Y_{i\ell}}\right]\right\}\nonumber \\
&\times \prod_{\ell=1}^L\left[{\sf TruncatedBeta}(\theta_\ell; a_\theta,b_\theta, (\max_{1\leq v \leq K^-_\ell}{\psi_{\ell v}}, 1)) \prod_{v}{\sf Beta}(\psi_{\ell v}; a_\psi,b_\psi)\mathbf{1}\{\bpsi_\ell \in \Delta\}\right] \cdot \nonumber\\
&\times  f(\alpha_1)\cdot {\sf IBP}_M({H}^*;\alpha_1, K)\cdot \PP(\cC; \gamma, p_K(\cdot)), \label{eq::joint}
\end{align} 
where $f(\alpha_1)$ is the density function of the hyperprior of  truncated IBP (to at most $M$ columns) parameter $\alpha_1$ and $\PP(\cC; \gamma, p_K(\cdot))$ is the prior in the space of partitions of observations.

\subsection{On Posterior Summary Given a Pre-specified Q}

In applications where $Q$ is known (Example 3), we infer for each subject the probability of having a latent state pattern $\bEta$, $\PP(\bEta_i = \bEta \mid \mathbf{Y})$, as estimated by the relative frequency of the event $\bEta_i = \bEta$ across MCMC iterations: $\frac{1}{B}\sum_{b=1}^B \mathbf{1}\{\bEta_i^{(b)}=\bEta\}, \forall \bEta \in \cA$ where $b$ indexes the stored MCMC samples obtained in Supplementary Material \ref{sec::split_merge}. Similarly, the posterior distribution for the total number of positive latent states $\PP(\sum_{m=1}^M\eta_{im}=z \mid \mathbf{Y})$ is estimated by the empirical frequencies $\frac{1}{B}\sum_{b=1}^B \mathbf{1}\{\sum_{m=1}^M\bEta_{im}^{(b)}=z\}$, $z=0, \ldots, M$, which in Example 3 represents the number of pathogens infecting the lung of a pneumonia child. To characterize the differential importance of each latent state among clusters, we also compute the posterior probability for $m$-th state being positive $\PP(\bEta^*_{(j)m}=1 \mid \{Y_i\})$, $j=1, \ldots, J'$, for $J'$ largest clusters across MCMC iteration.  Note that given $Q$, no merging or relabeling is required as in Step 3 and 7 in Supplementary Material \ref{sec::split_merge}. The number of scientific clusters $\tilde{K}$ can also be summarized by its empirical frequencies based on posterior samples.

\section{Marginal Likelihood $g(C)$}

To illustrate the calculation of marginal likelihood $g(C)$, we focus on two-parameter DINO model; see Remark \ref{remark::marginal_lkd_rlcm} for extensions to general restricted LCMs. Given assignment of subjects to clusters $\cC$, the model likelihood in a cluster $C_j\in \cC$ is
 \begin{align}
 pr\left(\{\bm{Y}_i, i\in C_j\} \mid \bEta^*_{j},\Theta,\Psi, Q \right) & = \prod_{\ell: \xi_{j\ell}=0} \psi_{\ell}^{n_{j\ell1}} \left(1-\psi_{\ell}\right)^{n_{j\ell0}}\cdot \prod_{\ell: \xi_{j\ell}=1} \theta_{\ell}^{n_{j\ell1}} \left(1-\theta_{\ell}\right)^{n_{j\ell0}},\label{eq::cluster_specific_likelihood}
 \end{align}
 where $n_{j\ell1} = \sum_{i: Z_i=j} Y_{i\ell}$ and $n_{j\ell0} = \sum_{i:  Z_i = j} (1-Y_{i\ell})$ are the number of positive and negative responses at dimension $\ell$ for subjects in cluster $C_j$, and $\xi_{j\ell} = \Gamma_{\bEta^*_j, \ell} = 1-\prod_{m=1}^M(1-\eta^*_{jm})^{Q_{m\ell}}$ indicates the true status for $\ell=1, \ldots, L$ and the product over $\ell$ is due to conditional independence given a cluster. We obtain the marginal likelihood $g(C)$ for cluster $C_j$ by integrating out latent states $\bEta^*_{j}$ in (\ref{eq::cluster_specific_likelihood}):
 \begin{align}
 g(C) & = \sum_{\bEta \in \{0,1\}^M}pr\left(\{\bm{Y}_i, i\in C_j\} \mid \bEta,\Theta,\Psi, Q \right)\PP(\bEta^*_j = \bEta\mid \bp),\label{eq::marg_lkd}
 \end{align}
 where $\PP(\bEta^*_j = \bEta\mid \bp) = \prod_{m=1}^M p_m^{\eta_m}(1-p_m)^{1-\eta_m}$. Note that $g(C)$ factorizes with respect to $\ell$ when $M=L$ and $Q=I_{L \times L}$ that leads to $\xi_{j\ell}=\eta^*_{j\ell}$. 
 
 \begin{remark}{\underline{Computational considerations.}}
 One of the computational costs results from the summation under a  large $M$ in (\ref{eq::marg_lkd}), or ``add" operation over $\bEta \in \{0,1\}^M$. The factorization with respect to $\ell$ allows the summations to be done for each $\ell$ separately and therefore reduces the number of ``add" operations  from $\cO(2^M)$ to $\cO(M)$  \citep[][Equation (8)]{hoff2005subset}. More generally, $g(C)$ also factorizes with respect to blocks that partition $\{1, \ldots, M\}$, $\{\cM_u, u=1, \ldots, U\}$ with $\cup {\cM_u}=\{1, \ldots, M\}$ when the corresponding row blocks of $Q$ are orthogonal ($\check{Q}_u = \lor_{m\in \cM_u}Q_{m\star}$, $u=1, \ldots, U$ are orthogonal), resulting in reduced ``add" operations $\cO(2^{\max_u |\cM_u|}L)$. Given $Q$,  we use Reverse Cuthill-McKee (RCM) algorithm \citep{cuthill1969reducing} for the $M$ by $M$ matrix $QQ^\top$ to simultaneously rearrange its rows and columns to obtain this block structure. 
 \end{remark}

\begin{remark}
\label{remark::marginal_lkd_rlcm}
To generalize (\ref{eq::cluster_specific_likelihood}) from two-parameter models to general restricted LCMs, simply replace the first product with $\prod_{\ell: \Gamma_{\bEta^*_j, \ell}=0}\left(\psi_{\ell, v(\bEta^*_i, \ell)}\right)^{n_{j\ell 1}}\left(1-\psi_{\ell, v(\bEta^*_i, \ell)}\right)^{n_{j\ell 0}}$.
\end{remark}

\section{Details of Posterior Algorithm}
\label{appendix_sec::details_posterior}
\subsection{Pre-specified Latent State Dimension $M < \infty$} 
\label{sec::split_merge}

When the number of components $K$ is unknown, one class of techniques updates component-specific parameters along with $K$. For example, the reversible-jump MCMC \citep[][RJ-MCMC]{green1995reversible} works by an update to $K$ along with proposed updates to the model parameters which together are then accepted or rejected. However, designing good proposals for high-dimensional component parameters can be non-trivial. Alternative approaches include direct sampling of $K$\citep[e.g.,][]{nobile2007bayesian, mccullagh2008many}. Here we build on the algorithm of \citet{miller2017mixture} for sampling clusters with discrete component parameters $\bEta^*_j$. We focus on model (6) in Main Paper to illustrate the posterior algorithm.

\begin{enumerate}

\item \underline{Initialization}. Initialize all model parameters from prior distributions. When a $Q_{m\star}$ is initialized to have redundant ones under high true positive rates, the likelihood of a sparse observation $\bY_i$ is much lower under $\eta_{im}=0$  than under $\eta_{im}=1$. Consequently, the sampling chain will visit $\eta_{im}=0$, i.e., inactive latent state $m$, with high probability. To better initialize active latent states, we therefore use a more stringent data-driven initialization for $Q_{\star \ell}$ by $Q_{m\ell} \overset{d}{\sim} {\sf Bernoulli}(p), m = 1, \ldots, M,$ only if many observations are positive at dimension $\ell$: $N^{-1}\sum_{i}Y_{i\ell} > \tau_1$, where $p$ and $\tau_1$ can be prespecified. In our simulations and data analysis, we set $p=0.1$ and $\tau_1=0.3$.

\item \underline{Split-merge update clusters $\cC$. }  

The one-subject-at-a-time, Gibbs-type update is typically slow in exploring a large space of clusterings. In fact, the number of ways to partition $N$ subjects is $B_N$, referred to as the Bell number and can be computed through the iterative formula $B_{N+1}= \sum_{n=0}^N {N \choose n} B_{n}$ with $B_0=B_1=1$ resulting in $B_{50}>2^{157}$. We remedy this by adding split-merge updates designed for conjugate models \citep[][]{jain2004split} that alter the cluster memberships for many subjects at once.

Because the Gibbs update (15) in Main Paper assigns clusters one subject at a time and updates clusters in a local fashion resulting in potential slow mixing of the sampling chain for $\cC$,  we use global updates to create or remove clusters for multiple subjects at a time that are likely to be accepted according to a Metropolis-Hastings ratio. We adapt an existing recipe designed for models with priors conjugate to the component-specific parameters \citep{jain2004split}, which uses split-merge updates to make global changes to cluster configuration followed by further refinement of clusters via Gibbs update one subject at a time. Given $\btheta$, $\bpsi$, $Q$ and $\mathbf{Y}$, a single split-merge update comprises the following steps:
\begin{enumerate}
\item[1a)] Randomly choose two observations $i$ and $j$ from $N$ subjects; Let $S$ be the indices of subjects either belonging to $C_{Z_i}$ or $C_{Z_{j}}$.
\item[1b)] Perform $r=5$ steps of \textit{intermediate} Gibbs scan (17) restricted to observations in the same clusters as $i$ or $j$. That is, use (17) to update observation $k\in S\setminus \{i,j\}$ with the constraint that $Z_k \in\{ Z_i,Z_j\}$; At the end of intermediate Gibbs scan, we obtain $\bm{Z}^{\sf \scriptsize launch}$. In this step, one assigns a subject $k$ in $S\setminus \{i,j\}$ to either the cluster of  $i$ or $j$ with probability 
\begin{align}
& \PP(Z_k = z \mid \bm{Z}_{-k}, \mathbf{Y}, {\sf~other~parameters~}) \nonumber\\
& =  \frac{(|C_{z}|+\gamma)g(C_{z} \cup \{k\})/g(C_{z})}{(|C_{Z_i}|+\gamma)g(C_{Z_i} \cup \{k\})/g(C_{Z_i})+(|C_{Z_j}|+\gamma)g(C_{Z_j} \cup \{k\})/g(C_{Z_j})}, z\in\{Z_i,Z_j\},\label{eq::restricted_gibbs}
\end{align}

\item[1c)] Perform a \textit{final} Gibbs scan restricted to observations  $S\setminus\{i,j\}$ using (\ref{eq::restricted_gibbs}) and obtain updated clusters as the proposal states to be used in a Metroplis-Hasting step which we denote  by $\bm{Z}^{\sf \scriptsize cand}$. We compute the proposal densities $q(\bm{Z}^{\sf \scriptsize cand} \mid \bm{Z})$ and $q( \bm{Z} \mid \bm{Z}^{\sf \scriptsize cand})$; For the non-trivial cases, the proposal densities depend on the random launch state $\bm{Z}^{\sf \scriptsize launch}$ and are products of Gibbs update densities in (\ref{eq::restricted_gibbs}).
\item[1d)] Accept or reject the proposed clustering $\bm{Z}^{\sf \scriptsize cand} $ with acceptance probability computed from prior ratio (based on two sets of clusters induced by $\bm{Z}^{\sf \scriptsize cand}$ vs $\bm{Z}^{\sf \scriptsize launch}$), likelihood ratio (given clusters $\bm{Z}^{\sf \scriptsize cand}$ vs $\bm{Z}^{\sf \scriptsize launch}$ and other population parameters), ratio of proposal densities (from 1c). See \citet{jain2004split} for the general recipe of computing the acceptance probability.
\item[1e)] Perform one complete Gibbs scan (17)  of $\bm{Z}$ for \textit{all} individuals to refine the current state of cluster indicators.
\end{enumerate}
The above is referred to as $(5,1,1)$ split-merge update where $5$ intermediate Gibbs scans are used to reach launch states $\bm{Z}^{\sf \scriptsize launch}$, one Metroplis-Hasting step to accept or reject a candidate clustering $\bm{Z}^{\sf \scriptsize cand}$, and one final complete Gibbs scan for all observations to refine the newly obtained cluster \citep{jain2004split}.


\item   Update individual machine usage profiles $H=\{{\eta}_{im}\}$. Because subjects within a cluster share latent states $\bEta_i = \bEta^*_j$, $i\in \{i: Z_i = j\}$ for cluster $j=1, \ldots, T$, we sample from
\begin{align}
[{\bEta}^*_{j} \mid {\sf others}]  \propto &~    \prod_{m=1}^M\{p_m\}^{\eta^*_{jm}}\{1-p_m\}^{1-\eta^*_{jm}}\cdot\prod_{\ell: \xi_{j\ell}=0} \psi_{\ell}^{n_{j\ell1}} \left(1-\psi_{\ell}\right)^{n_{j\ell0}}\cdot \prod_{\ell: \xi_{j\ell}=1} \theta_{\ell}^{n_{j\ell1}} \left(1-\theta_{\ell}\right)^{n_{j\ell0}},\nonumber
\end{align}
where $\xi_{j\ell}=\Gamma_{\bEta^*_j, \ell}$ indicates the active or inactive status at dimension $\ell$ in cluster $C_j$, $\bp=\{p_m\}$ are within-cluster prevalence of $M$ latent states and $n_{j\ell1}=\sum_{i: Z_i = j}Y_{i\ell}$ and $n_{j\ell0}=\sum_{i: Z_i = j}(1-Y_{i\ell})$. Because $\bEta_j^{*}\in\{0,1\}^M$, it is important to move around in this space fast. We currently use multinomial sampling in simplex $\Delta^{2^M-1}$, which can be improved by either Hamming ball sampler or parallel tempering.

We remark on ``partner latent states" that motivate merging a subset of rows in $Q^{(b)}$. Let $H^{(b)} = \{\eta^{(b)}_{im}\}$ be an $N$ by $M$ binary matrix that collects latent states for all subjects at iteration $t$. Let $M^{(b)}_{\sf eff} = \sum_{m=1}^{M} \ind\{\mathbf{1}^\top_N H_{\star m}^{(b)}\neq 0\}$ be the number of nonzero columns in ${H}$ at $t$-th MCMC iteration. The identifiability conditions apply only to the first $M^{(b)}_{\sf eff}$ rows of $Q$. Condition (C1) and (C3) hold at each iteration regardless of the value of $M^{(b)}_{\sf eff}$ because $Q\in \mathcal{Q}$ truncated to first $M^{(b)}_{\sf eff}$ rows remains in $\mathcal{Q}$. At each iteration, conditions (C1) and (C3) also hold if we collapse two identical columns $(m,m')$ of ${H}^{(b)}$ to combine two partner machines that are present or absent together among subjects ($\eta_{im}^{(b)} = \eta_{im'}^{(b)}$, $i=1, \ldots, N$); We set $H^{(b)}_{\star m'}= \mathbf{0}_{N}$ and the other row ${Q}^{(b)}_{m\ell}= \max\{Q^{(b)}_{m\ell},Q^{(b)}_{m'\ell}\}$, $\ell = 1, \ldots, L$. It is easy to verify that this scheme preserves conditions (C1) and (C3) and readily generalizes to cases where more than two columns of $H^{(b)}$ are identical. In the population, the diversity assumption $\cA=\{0,1\}^M$ does not hold if two latent states always positive together. When external knowledge is available for two ``partner" states with separate known rows in $Q$, it can be readily integrated into posterior sampling.

\item Sample false positive rates from \[
[\psi_\ell \mid {\sf others} ] ~{\sim} ~ {\sf Beta}\left(\sum_{i}(1-\xi_{i\ell})Y_{i\ell}+a_\psi,
\sum_{i}(1-\xi_{i\ell})(1-Y_{i\ell})+b_\psi\right)\ind\{(0,\theta_\ell)\}, \ell = 1, \ldots, L.\]

Sample true positive rates from
\[
[\theta_\ell \mid {\sf others} ]~{\sim}  ~{\sf Beta}\left(\sum_{i}{\xi_{i\ell}Y_{i\ell}}+a_\theta,
\sum_{i}\xi_{i\ell}(1-Y_{i\ell})+b_\theta\right)\ind\{(\psi_\ell,1)\}, \ell =1,\ldots, L.
\]
We also implemented in ``\verb"rewind"" specified upper bounds for $\{\psi_\ell\}$ and lower bounds for $\{\theta_\ell\}$ when needed.

\item Update hyperparameter $\alpha$. Suppose the hyperprior for $\alpha$ is $p(\alpha)$. Then by the marginal distribution of $H^*$ from finite-$M$ IBP \citep{ghahramani2006infinite}, we reparametrize in terms of $\beta = \frac{\alpha}{\alpha+1} \in (0,1)$ and obtain 
\[[\beta\mid H^*] \propto p(\beta) \cdot \left(\frac{\beta}{1-\beta}\right)^M  \prod_{m=1}^M\frac{\Gamma(s_m+\beta/\{M(1-\beta)\})}{\Gamma(T+1+\beta/\{M(1-\beta)\}))},\]
which can be sampled from a dense grid over $(0,1)$ and $s_m = \sum_{j=1}^T \eta^*_{jm}$ is the number of clusters that $m$-th latent state is positive. We use Beta distribution $\beta \sim {\sf Beta}(a_\beta,b_\beta)$ where $a_\beta=b_\beta=1$ in our simulations and data analyses.

\item Update prevalence parameters $\bp=\{p_1, \ldots, p_m\}$ from
\begin{align}
[\bm{p}\mid {\sf others}] & \propto ~ \prod_{m=1}^M(p_m)^{n^*_{m1}}(1-p_m)^{n^*_{m0}} {\sf Beta}(p_m; \alpha/M,1),
\end{align}
which we sample independently $p_m\sim {\sf Beta}(n^*_{m1}+\alpha/M, n^*_{m0}+1)$, $m=1, \ldots, M$.


\item Update machine  matrix $Q$ via constrained Gibbs sampler. Update to $Q_{m\ell}^{(b)}$, $\ell = 1, 2, \ldots, L$, $m=1, 2, \ldots, M$ under two mutually exclusive scenarios:
\begin{enumerate}
\item[1a)] Keep $Q_{m\ell}^{(t-1)}$ if one of the three criteria holds: 1) $Q^{(t-1)}_{\star\ell} = \bm{e}_m$, 2) ${1}_{L}^\top Q^{(t-1)}_{m,\star}=3$ and $Q_{m\ell}=1$ or 3) $Q_{m\ell}^{(t-1)}=0$, $Q_{\star\ell}^{(t-1)}=\bm{e}_m$ and there are only two $\bm{e}_m$ in the columns of $Q$.
\item[1b)] Otherwise, flip $Q_{m\ell}^{(t-1)}$ to a different value $z$ with probability $p(z\mid {\sf others})/(1-p(z\mid {\sf others}))$, where $p(z\mid {\sf others})$ is the full conditional distribution 
\begin{eqnarray}
pr(Q_{m\ell}=z \mid {\sf others}) &\propto & \prod_{i=1}^N pr\left(Y_{i\ell}\mid \{\bm{\eta}_i\}, Q^{(b)}_{\sf new},Q_{m\ell}=z, Q_{\sf old}^{(t-1)}, \theta_\ell, \psi_\ell\right)\nonumber \\
& = & \prod_{i: \xi_{i\ell}=1} \theta_\ell^{n'_{1\ell 1}}(1-\theta_\ell)^{n'_{1\ell 0}}\cdot  \prod_{i: \xi_{i\ell}=0} \psi_\ell^{n'_{0\ell 1}}(1-\psi_\ell)^{n'_{0\ell 0}}, z=0,1, \nonumber 
\end{eqnarray}
where $n'_{1\ell1} = \sum_{i=1}^N \xi_{i\ell}Y_{i\ell}$,  $n'_{1\ell0} = \sum_{i=1}^N \xi_{i\ell}(1-Y_{i\ell})$, $n'_{0\ell1} = \sum_{i=1}^N (1-\xi_{i\ell})Y_{i\ell}$, $n'_{0\ell0} = \sum_{i=1}^N (1-\xi_{i\ell})(1-Y_{i\ell})$, and $Q^{(b)}_{\sf new}$ and $Q_{\sf old}^{(t-1)}$ represent entries of $Q$ that have and have not been updated, respectively. 

\item[2)]Permute the rows of $Q^{(b)}$ by natural ordering of binary codes $\{Q_{m\star}, m=1, \ldots, M\}$ represented in binary system. We order the rows of $Q^{(b)}$ by decreasing order of $M$-dimensional vector $Q^{(b)}\bm{v}$ where $\bm{v} = (2^{L-1},2^{L-2}, \ldots, 1)^\top$. We only do so after all the MCMC iterations.

Condition (C1) guarantees that once $Q^\top$ is written in left-ordered form \citep{ghahramani2006infinite}, the bottom row of ${Q}$ corresponds to a row with a positive \textit{ideal} response at the smallest dimension $\ell_{(1)}=\arg\min_\ell \{Q_{m\ell}=1, \forall m,\ell\}$, which if shared by more than one row, then the row having a postive ideal response at the second lowest dimension  $\ell_{(2)}=\arg\min_{\ell:\ell > \ell_{(1)}} \{Q_{m\ell}=1, \forall m,\ell\}$ is placed at the bottom row; this scheme of ordering the rows of $Q$ will always succeed according to (C1).  
\end{enumerate} 

Finally, suppose at iteration $s$, the MCMC algorithm produces latent states unused by any observation: $\cM^{\sf non,(b)}=\{m': \sum_i \eta^{(b)}_{im'} =0\}$.  We reset to zeros the subset of rows of $Q$ corresponding to the unused latent states at an iteration. Given the sampled $\bEta^{(b)}_{i}$, the corresponding set of rows $Q^{(b)}_{\cM^{\sf non}}=\{Q^{(b)}_{m\star}, m\in \cM^{\sf non,(b)}\}$ does not enter likelihood. We re-initiate $Q^{(b)}_{\cM^{\sf non}}$ which upon sequential Gibbs scans create new machines that may enter and improve the likelihood at the next iteration. In our experiments, resetting $Q^{(b)}_{m\star}$ side-steps the difficulty of splitting a sampled machine that is populated with too many ones. Resetting is also practically easier to implement compared to a fine-tuned split-merge algorithm applied to the rows of $Q$ in tandem with simulated annealing which are designed for a more complex time series segmentation tasks  \citep[e.g.,][]{fox2014joint}. 

\noindent \textit{Convergence checks.} In simulations and data analysis, we ran three MCMC chains each with a burn-in period of $10,000$ iterations followed by $10,000$ iterations stored for posterior inference. We look for potential non-convergence in terms of Gelman-Rubin statistic \citep{brooks1998general} that compares between-chain and within-chain variances for each model parameter where a large difference ($R_c>1.1$) indicates non-convergence; We also used Geweke's diagnostic \citep{geweke1996measuring} that compare the observed mean for each unknown variable using the first $10\%$ and the last $50\%$ of the stored samples where a large $Z$-score indicates non-convergence ($|Z|>2$). In our simulations and data analyses, we observed fast convergence (many satisfied convergence criteria within $2,000$ iterations) that led to well recovered clusters and $Q$ matrices (results not shown here).

\end{enumerate} 

\subsection{Algorithm under $M=\infty$}
\label{sec::unknown_M}
This section presents the algorithm without the need to pre-specify the exact or an upper bound of the number of factors $M$. The algorithm adapts the slice sampler for infinite factor model \citep{teh2007stick} which performs adaptive truncation of the infinite model to finite dimensions and avoids approximation of the Indian Buffet Process (IBP) prior for $H^*$. The algorithm builds on the \textit{semi-ordered} representation of the IBP, where the probabilities of active states are \textit{non-ordered} and the probabilities of inactive states truncated to a random number $M^0$ are \textit{ordered}. We use this algorithm to infer the number of active states.

\begin{itemize}
\item[0.] Initialize the number of active states $M^+$, the random truncation level for inactive states $M^0=0$. Initialize $Q$ with an appropriate $M^*=M^++M^0$ by $L$ binary matrix; Initialize the IBP hyperparameter $\alpha$; Initialize $p$ of length $M^*$ to be the vector of the probabilities for each state being used (if the initial $M^0=0$ as recommended, then $p$ needs not be ordered). Initiate $H^*$ as $(T_{\max}+3)$ by $M_{\max}$ matrix with all zeros, where $T_{\max}$ and $M_{\max}$ are the guessed maximum number of clusters and truncated number of states the algorithm will visit across iterations. Neither $T_{\max}$ nor $M_{\max}$ is introduced to approximate any probabilistic distribution: one can increase both numbers as appropriate at the expense of extra memory.  

\noindent \underline{Repeat steps 1 to 10 below for iterations $b=1, \ldots, B$:}
\item[1.] Gibbs update cluster indicators $\bZ=\{Z_i, i=1, \ldots, N\}$ and the cluster-specific sizes $|\cC_j|, j=1, \ldots, t$, where $t$ is the number of unique values in $\bZ$
\item[2.] For Iteration 1, update $H^*$ elementwise for $t\cdot M^*$ elements corresponding to the currently non-empty clusters and the current truncation level $M^*$ for the number of factors; Otherwise, update $H^*$ by the full conditional distribution given other parameters including the slice variable $s$:
\begin{align}
& pr(\eta^*_{jm} = z \mid {\sf others}) \propto \frac{p_m}{p^+_{\min}}\times\nonumber\\
& \prod_{m=1}^M\{p_m\}^{\eta^*_{jm}}\{1-p_m\}^{1-\eta^*_{jm}}\cdot\prod_{\ell: \xi_{j\ell}=0} \psi_{\ell}^{n_{j\ell1}} \left(1-\psi_{\ell}\right)^{n_{j\ell0}}\cdot \prod_{\ell: \xi_{c\ell}=1} \theta_{\ell}^{n_{j\ell1}} \left(1-\theta_{\ell}\right)^{n_{j\ell0}},\nonumber
\end{align}
for $z=0,1$, $m=1, \ldots, M^+$, where $p^+_{\min}=p^+_{\min}(H^*, \{p_{m},m=1, 2, \ldots, \})=\min_{1\leq m \leq M^+}\{p_m\}$ depends on $\eta^*_{jm}$ and is the normalizing constant for the uniform distribution of the slice variable: $pr(s \mid H^*,\{p_{m},m=1,2,\ldots,\})=\mathbf{1}_{\{0\leq s \leq p^+_{\min}\}}/p^+_{\min}$. For example, given $s$ one must set to zero any column $m\in \{1,\ldots, M^+\}$ in $H^*$, $\{\eta^*_{jm},j=1, \ldots, t\}$ whenever $p_m < s$.
\item[3.] Update $Q$ matrix ($M^*$ by $L$) as in Step 6 in Section \ref{sec::split_merge};
\item[4.] Update the number of active factors ($M^+$) by finding the number of columns in $H^*$ with non-zero column sums.
\item[5.] Update unordered $\{p_m,m=1, \ldots, M^+\}$ by $p_m\sim {\sf Beta}(\sum_{j=1}^t \eta^*_{jm}, 1+t-\sum_{j=1}^t \eta^*_{jm})$, $m=1, \ldots, M^+$;
\item[6.] Update slice variable $s\sim {\sf Uniform}(0,\min_m p_m)$;
\item[7.] Starting from $m=1$, sample 
\[p^0_{(m)} \mid p^0_{(m-1)} \sim \exp\{\alpha\sum_{j=1}^t(1-p_{(m)}^0)^j\}(p^0_{(m)})^\alpha(1-p^0_{(m)})^N\cdot \mathbf{1}_{\{0\leq p^0_{(m)}\leq p^0_{(m-1)}\}},\] 
until $p^0_{(M^0+1)}<s$, where $p^0_{(0)}=1$. Use adaptive rejection sampling \citep[ARS,][]{gilks1992adaptive} to sample from this distribution iteratively for $m=1, \ldots, M^0$, where $M^0>0$ only when $p^0_{(1)} > s$;
\item[8.] If $M^0>0$, update $p$ by concatenating the old $p$ and $p^0$; update $M^*=M^++M^0$;
\item[9.] Pad $H^*$ with $M^0$ columns of zeros to its right; Subset the rows of $Q$ to those $M^+$ factors and pad it with $M^0$ extra rows sampled from an appropriate initialization sampler;
\item[10.] Update other parameters $\btheta$, $\bpsi$, $\alpha$ as in Section \ref{sec::split_merge}.
\end{itemize}

\section{Likelihood-based identifiability conditions given $\tilde{K}$, $M$ and $\Gamma$ (or $Q$)}
\label{appendix::ident_condition}
Given $\Gamma$, \citet{gu2018} established that the separability of $\Gamma$ is sufficient and necessary for identifying $\bpi_{\tilde{K}}$ under two-parameter models for \textit{known} conditional response probabilities $\Lambda$; If $\Gamma$ is inseparable, $\bpi_{\tilde{K}}$ is identified up to equivalent classes defined by identical rows in $\Gamma$ (in this paper, we transposed $\Gamma$ used in \citet{gu2018}). When $\Lambda$ is unknown, \citet{gu2018} established sufficient conditions for $\bpi_{\tilde{K}}$-partial identifiability (strictly identify $\Lambda$ but identify $\bpi_{\tilde{K}}$ up to equivalent classes defined by identical rows in $\Gamma$). For $Q$-restricted two-parameter models, if $\cA$ is saturated and $\Gamma$ is separable, then these conditions become  minimal, i.e.  sufficient and necessary conditions: 1) $\geq 3$ items per latent state and 2) $Q=[I_M,Q_1^\top]$ where $Q_1$ has distinct columns. 

For multi-parameter models, separability of $\Gamma$ is sufficient for identifying $\bpi_{\tilde{K}}$ given known $\Lambda$. $\bpi_{\tilde{K}}$ will be strictly identifiable  given two technical conditions \citep[][C3 and C4]{gu2018} - Condition (C3) implies separability of $\Gamma$ which could be true for $Q$-RLCM induced $\Gamma$ with unsaturated $\cA$ and without single-attribute items in $Q$. They also established ``generic identifiability" results for $\Lambda$ and $\bpi_{\tilde{K}}$ when $\Gamma$ is inseparable: as long as one can flip entries to satisfy two technical conditions. The notion of ``generic identifiability" is introduced, because the identifiability results for multi-parameter models hold except on a Lebesgue measure-zero set where the models are reduced to two-parameter models. For the special cases of $Q$-restricted model (saturated), the two technical conditions do not require the $Q$-matrix to contain an identity submatrix and provides a flexible new condition for generic identifiability under various $Q$-matrix structures; the results are generically identifiable up to label swapping among those latent classes that have the same row vectors in the $\Gamma$-matrix.

\section{Additional simulated example: removing irrelevant features reduces the noise and improves cluster estimation}
\label{appendix::simulation_feature_selection}

When $Q$ is unknown, the proposed method for scientifically structured clustering includes an additional step for sampling $Q$. A zero column in $Q$,  say column $\ell$,  indicates irrelevance of $\ell$-th dimension because all positive observations at that dimension will be false positives. By estimating which columns are zeros, our algorithm removes irrelevant features when clustering observations. 

Clustering multivariate binary data on a subset of features reduces the impact of noise introduced by less important features and therefore can be superior to all-feature clustering methods such as the standard latent class analysis. For example, in model (2) with $Q = I_{L\times L}$, irrelevant features $ \cL^c=\{\ell: \Gamma_{\star\ell}=\bm{0}\}$ ideally would not enter likelihood ratio calculations when assigning observations to clusters. Indeed, let $R_{kk'}(\bY=\by)$ be the log relative probabilities of assigning an observation $\bY_i$ to cluster $k$ ($\cC^{(k)}_{-i}$) versus $k'$ ($\cC^{(k')}_{-i}$) given other parameters and clustering $\cC_{-i}$ can be Taylor approximated by  
\begin{align}
R_{kk'}(\bY_i) & \approx \log \frac{|\cC^{(k)}_{-i}|+\gamma}{|\cC^{(k')}_{-i}|+\gamma}+\sum_{\ell=1}^L p_\ell \log \left(\frac{\hat{\theta}_{(k)\ell}}{\hat{\theta}_{(k')\ell}}\right)^{Y_{i\ell}}  \left(\frac{1-\hat{\theta}_{(k)\ell}}{1-\hat{\theta}_{(k')\ell}}\right)^{1-Y_{i\ell}},\label{eq::relative_prob}
\end{align}
where the terms corresponding to irrelevant features become negligible if $\hat{\theta}_{(k)\ell} \approx \psi_{\ell}$. The response probabilities at irrelevant dimensions ($\{\psi_\ell: \ell\in \cL^c\}$) are nevertheless estimated with error and contribute to noise in assigning each observation to an existing cluster. $R_{kk'}(\bY) >0, =0, <0$ indicate assignment of observation $\bY$ to cluster $k$ more, equally and less likely than to cluster $k'$, respectively. Consider a triple of observations ($\bY_1, \bY_2, \bY_3$) where the first  (cluster $k'$) and the rest (cluster $k$) belong to two distinct clusters, respectively. The probability of clustering $\bY_1$ and $\bY_2$ into their respective true clusters is $p_{12} = (1-{\sf expit}\{R_{kk'}(\by_1)\}){\sf expit}\{R_{kk'}(\by_2)\}$; the probability of assigning $\bY_2$ and $\bY_3$ into the same true cluster is $p_{23}={\sf expit}\{R_{kk'}(\by_2)\}{\sf expit}\{R_{kk'}(\by_3)\}$. Here we have used lower case $\by_i$ to represent the sub-vector of $\bY_i$ that entered the calculation in (\ref{eq::relative_prob}).

We simulated $L_1=5$ relevant dimensions and $L_2=40$ irrelevant dimensions $\cL^c = \{6, \ldots, 45\}$. To mimic the noisy estimates of the response probabilities in cluster $k$ and $k'$, we simulated $\hat{\theta}_{(k)\ell} = (\log \br, \log \bepsilon)$ and $\hat{\theta}_{(k')\ell} = (\log \br', \log \bepsilon')$ where $r_{\ell1}, \ldots, r_{\ell, L_1}\overset{d}{\sim}{\sf Beta}(0.1N_k,0.9N_k)$, $r'_{\ell1}, \ldots, r'_{\ell, L_1}\overset{d}{\sim}{\sf Beta}(0.9N_{k'},0.1N_{k'})$ and $\epsilon_{\ell1}, \ldots, r_{\ell, L_2}\overset{\sf iid}{\sim}{\sf Beta}(0.1N_k,0.9N_k)$ and $\epsilon'_{\ell1}, \ldots, \epsilon'_{\ell, L_2}\overset{\sf iid}{\sim}{\sf Beta}(0.1N_{k'},0.9N_{k'})$. We set $N_k = N_{k'}=20$. Given $\{\hat{\theta}_{(k)\ell}\}$ and $\{\hat{\theta}_{(k')\ell}\}$, we draw observations from two classes that have response probability profiles ($\bY_2$ and $\bY_3$ from $\{\theta_{(k)\ell}, \ell=1, \ldots, L\}=(\underbrace{0.9,\ldots, 0.9}_{L_1}, \underbrace{0.1, \ldots, 0.1}_{L_2})$ and $\bY_1$ from $\{\theta_{(k')\ell}, \ell=1, \ldots, L\}=(\underbrace{0.1,\ldots, 0.1}_{L_1}, \underbrace{0.1, \ldots, 0.1}_{L_2})$).

Based on $R=100$ replications, Figure \ref{fig::shrinkage} shows $R=100$ values of $p_{12}$ (left) and $R=100$ values of $p_{23}$ (right) computed by setting $\{\by_i, i=1, 2,3\}$ to be the irrelevant, all  and relevant features in the data vector $\{\bY_i,i=1,2,3\}$, respectively. 

By selecting relevant features, the model improves our ability to separate observations from distinct clusters and group observations that belong to the same cluster. On the left panel, the all-feature $p_{12}$ values are pulled towards zero (towards left) that favors assigning $\bY_2$ to cluster $k$ and $\bY_1$ to cluster $k'$. On the right panel, the all-feature $p_{23}$ values are pulled towards one (towards right) that favors clustering $\bY_2$ and $\bY_3$ together in the true cluster ($k$). 

In practice, the relevant features are of course to be inferred from data, by their observed marginal independence from the rest of the measured features. The improvements of clustering using subset clustering with \textit{inferred} subsets can be seen from in Figure 2 in Main Paper by the superior clustering performance in (f) under feature selection compared to (e) obtained without selecting features.

\begin{figure}[H]
\captionsetup{width=0.95\linewidth}
\centering
\addtocounter{figure}{1} 
\includegraphics[width=\linewidth]{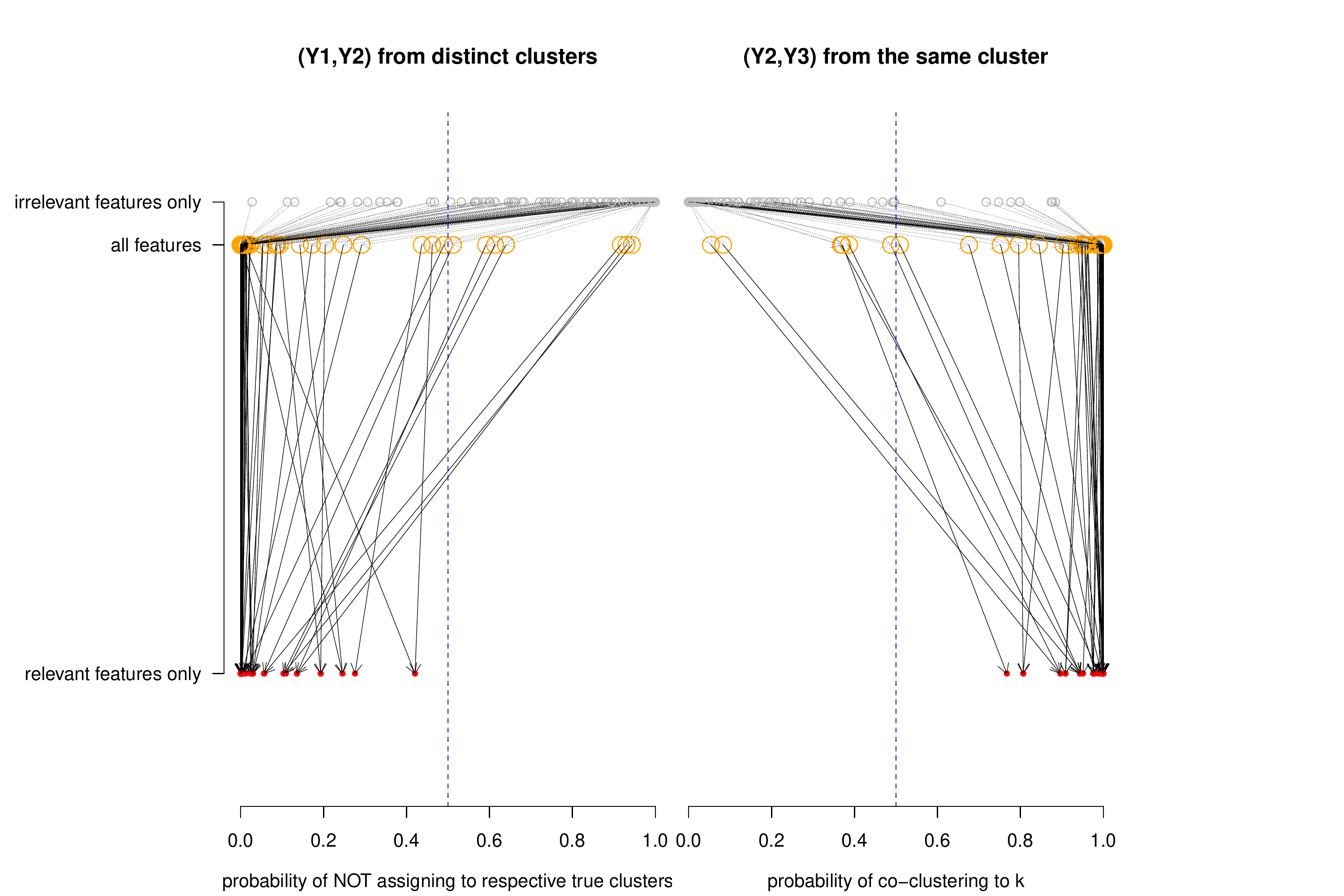}
\label{fig:truth}
\addtocounter{figure}{-1} 
\caption{Removing irrelevant features improves estimation of clusters. \textit{Left}) 100 random pairs of observations drawn from distinct clusters; the probability of them not being clustered correctly is lowered (pulled towards zero) once the irrelevant features are removed. \textit{Right}) 100 random pairs of observations drawn from the same cluster; the probability of co-clustering to the correct cluster is increased towards one once the irrelevant features are removed. }
\label{fig::shrinkage}
\end{figure}

\subsection{Additional Figures and Tables}
\begin{landscape}
\begin{table}[]
\begin{threeparttable}
\centering
\label{my-label}
\begin{tabular}{ccc|cc|cc}
\multicolumn{3}{c}{}& \multicolumn{4}{|c}{Methods (examples)}\\ \cline{1-7}
\multicolumn{3}{c|}{\huge \shortstack{}} &\multicolumn{2}{c|}{Restricted LCM}  & \multicolumn{2}{c}{LCM}  \\
\multicolumn{3}{c|}{\bf \shortstack{Model Specification}} &\multicolumn{1}{c}{\shortstack{Bayesian}} & \multicolumn{1}{c|}{\shortstack{non-Bayesian}} & \multicolumn{1}{c}{\shortstack{~Classical}} & \multicolumn{1}{c}{\shortstack{Nested Partially}$^\dagger$}  \\[0.2cm] \hline 
\multirow{3}{*}{{\bf\footnotesize \shortstack{ latent state \\ variables \\($\bEta_i\in \cA\subset \{0,1\}^M$;\\$\#$latent classes:\\$\tilde{K}=\mid \cA \mid$)}}} & \multirow{1}{*}{\footnotesize \shortstack{$\tilde{K}$ known}}& \shortstack{\footnotesize\\~\\~$\cA$ pre-specified \\~} &{\scriptsize \shortstack{$\cA=\{0,1\}^M$:\\\citet{chen2017bayesian}}} &{\scriptsize \shortstack{$\cA = \{0,1\}^M$:\\~ \citet{xu2017identifiability};\\$\mathbf{0}_M\in \cA \subsetneqq \{0,1\}^M$:\\~\citet{leighton2004attribute},\\\citet{gu2018}}} &  {\scriptsize \shortstack{\\~\citet{lazarsfeld1950thelogical}$^\star$,\\
\citet{anderson1954estimation}$^\star$,\\\citet{Goodman1974}$^\star$\\~\citet{erosheva2007describing}$^{\dagger,\ddagger}$,\\
\citet{garrett2000latent}$^{\dagger}$}} & {\scriptsize\shortstack{$\mathbf{0}_M \in \cA$ and \\$ \text{~partially~observed~}$\\~ some of $\{i: \bEta_i = \mathbf{0}_M\}$: \\ \citet{wu2017nplcm} }} \\ \cline{3-3}
&& \shortstack{\footnotesize \\~\\~$\cA$ unknown\\~\\~} & \shortstack{{\scriptsize(proposed)}}  & {\scriptsize \shortstack{ \citet{miettinen2008discrete}$^\#$\\ ($Q=I_{L\times L}$) }}& -&  - \\ \cline{2-7}
&\multirow{1}{*}{\shortstack{\footnotesize $\tilde{K}$ unknown}}&\shortstack{\footnotesize \\~$\cA$ unknown}   &\shortstack{{\scriptsize(proposed)}} & - &\shortstack{\scriptsize\citet{dunson2009nonparametric}$^\dagger$}&   \shortstack{\scriptsize \citet{hoff2005subset} }   \\ 
&&&&&&\\ \cline{1-7}
\multirow{2}{*}{{\bf \footnotesize \shortstack{design matrix \\($\Gamma = \left(\Gamma_{\bEta,\ell}\right)$\\$ \in\{0,1\}^{\tilde{K}\times L}$)}}}&\multirow{2}{*}{\shortstack{\footnotesize\\~\\~\\~\\~\\~\\~\\~\\\\~$Q$-matrix\\($\Gamma=\Gamma(\bEta,Q)$)}}& \shortstack{\\~\\~\\~known} &  {\scriptsize \shortstack{(proposed)}}  & {\scriptsize\citet{xu2017identifiability} }  & \checkmark: \multirow{1}{*}{$Q=\mathbf{1}_{M \times L}$} & {\scriptsize \shortstack{ \scriptsize\\~ \citet{wu2017nplcm}; \\ \citet{hoff2005subset}: $Q=I_{L\times L}$}}  \\ \cline{3-3}
& &\shortstack{\\~\\~\\~unknown} &  {\scriptsize \shortstack{(proposed),\\\citet{chen2017bayesian},\\\citet{rukat2017bayesian}}} &  {\scriptsize \shortstack{\citet{XuShang2017},\\
\citet{chen2015statistical}} }& - &  - \\ [0.4cm]\cline{1-7}
\multirow{6}{*}{{\bf \footnotesize\shortstack{\footnotesize\\~\\~\\~\\~\\~\\~\\~\\~\\~\\~\\~\\~\\~\\~\\~\\~\\~\\~\\~\\~\\~\\~measurement\\ process\\($[\bY_i \mid \bEta_i, \Gamma, \bm{\Lambda}]$)\\~}}}
&\multirow{2}{*}{\shortstack{\footnotesize\\local indep.\\given $\bEta_i$~}}& yes & \shortstack{{\scriptsize(proposed)}}  & \checkmark &  \checkmark &   {\scriptsize\citet{wu2016partially} }\\\cline{3-3}
&& \shortstack{\\no\\~} & -  & - & {\scriptsize \shortstack{\citet{pepe2006insights},\\\citet{albert2001latent}} }  &   {\scriptsize\citet{wu2017nplcm} }  \\
\cline{2-7}
&\multirow{4}{*}{\shortstack{\footnotesize\\~\\~\\~\\~\\~\\~\\~\\~\\~$(K^+_\ell,K^-_\ell)$\\~}}& \shortstack{\\$(=1,=1)$\\~} &{\scriptsize \shortstack{(proposed),\\\citet{chen2017bayesian},\\\citet{rukat2017bayesian},\\\citet{wu2016partially}}}  &  {\scriptsize \shortstack{\\\citet{junker2001cognitive},\\\citet{templin2006measurement}}}  &  - &   {\scriptsize \shortstack{\citet{wu2016partially}}}  \\ \cline{3-3}
&& \shortstack{\\$(\geq 1,=1)$\\~} & -  &  - &-  &   {\scriptsize \shortstack{\citet{hoff2005subset}}}  \\ \cline{3-3}
&& \shortstack{\\$(=1,\geq 1)$\\~}  & \shortstack{{\scriptsize (proposed)}}  &  {\scriptsize \shortstack{\\~\\~\\\citet{de2011generalized},\\\citet{henson2009defining}}}  & -  &  - \\ \cline{3-3}
&& \shortstack{\\$(\geq 1,\geq 1)$\\~} & -  &  - &  - & \multirow{1}{*}{{\scriptsize \shortstack{\citet{wu2017nplcm}}}}  \\\cline{3-3}
&& \shortstack{\\$(\geq 1,=0)$\\~} & -  &  - &  \checkmark & - \\\cline{1-7}
\end{tabular}
\begin{tablenotes}
\item $^\dagger$: Bayesian approach.
\item $^\ddagger$: has equivalent LCM formulation.
\item $^\#$: non-probabilistic
\item $^\star$: early applications.
\item \checkmark: applies to all in the column (except for other rows in the same row block)
\end{tablenotes}
\caption{Comparison of variants of latent class analysis of multivariate binary data.}
\label{table::graph_biblio}
\end{threeparttable}
\end{table}
\end{landscape}



\begin{landscape}
\begin{figure}[h]
\captionsetup{width=.9\linewidth}
\centering

\includegraphics[width=\linewidth]{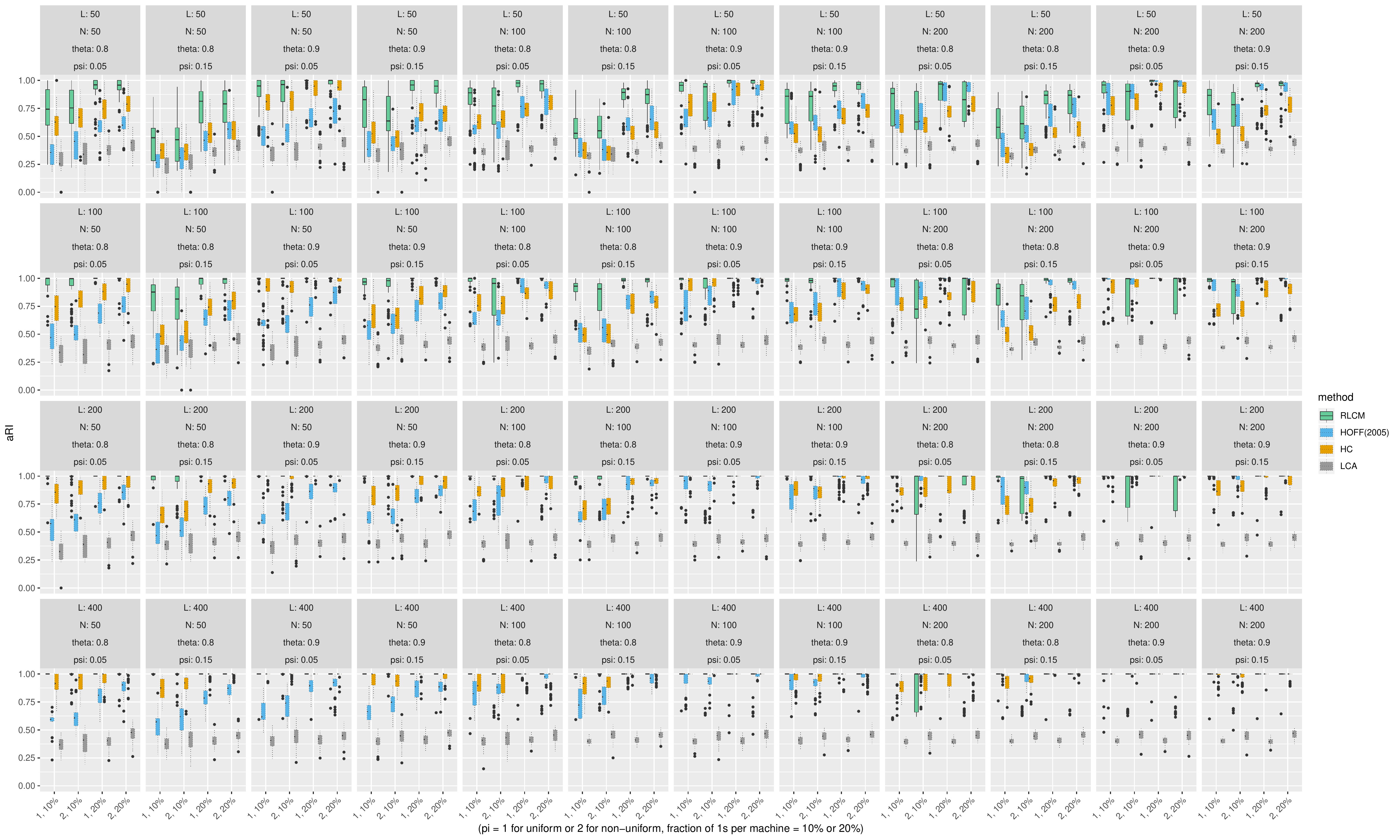}
\label{fig:simulation_replication}

\caption{Based on $R=60$ replications for each parameter setting, Bayesian {\sf RLCM} (boxplots with solid lines) most accurately recovers the true clusters compared to subset clustering (Hoff, 2005) hierarchical clustering ({\sf HC}) and traditional Bayesian latent class analysis ({\sf LCA}) (from the left to the right in each group of four boxplots). This figure expands Figure 2 in Main Paper over more parameter settings.}
\label{fig:simulation_replication}
\end{figure}
\end{landscape}

\begin{figure}[h]
\captionsetup{width=0.8\linewidth}
\centering
\addtocounter{figure}{1} 
\includegraphics[width=0.8\linewidth]{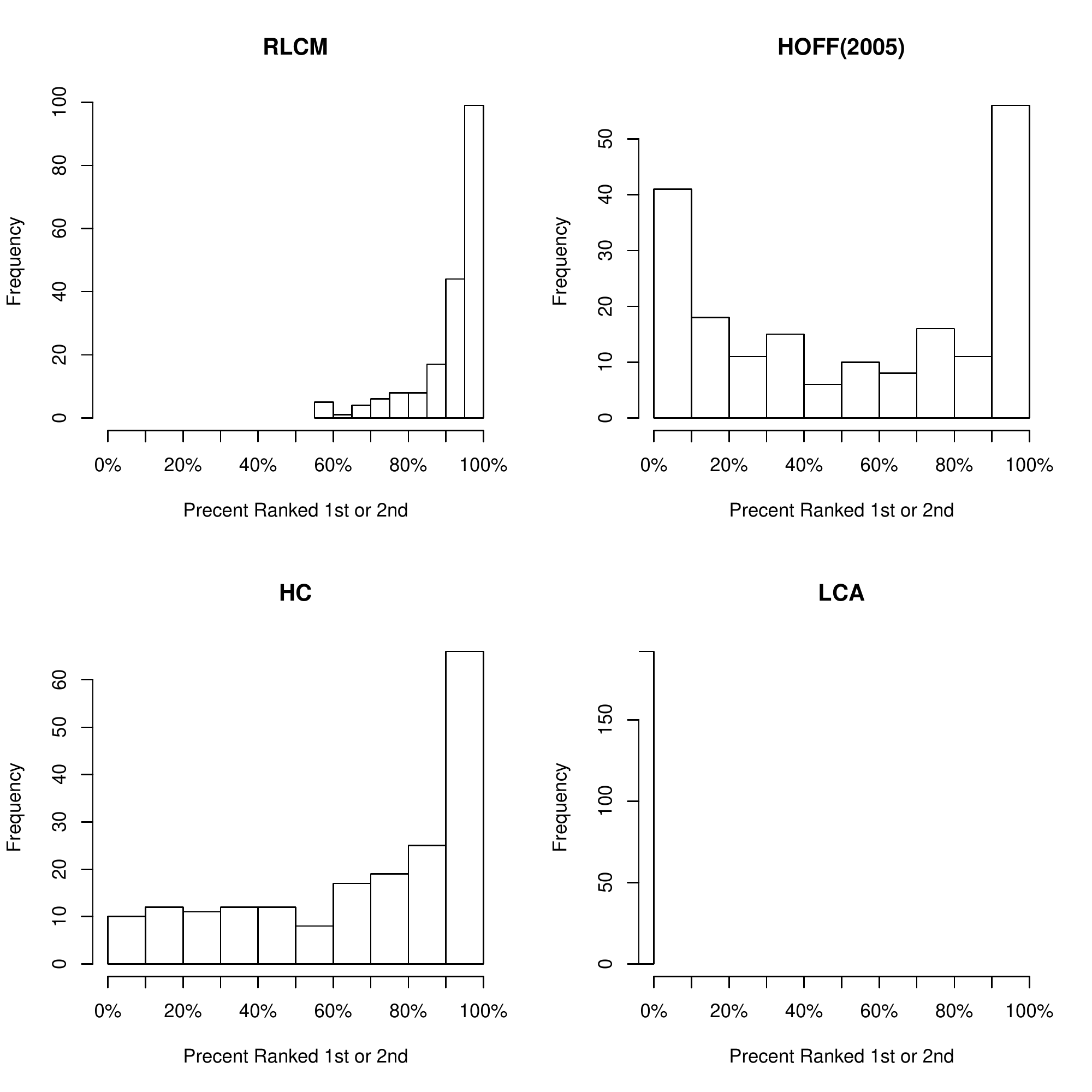}
\label{fig:hist_first_second}
\addtocounter{figure}{-1} 
\caption{For each of four clustering methods (Bayesian {\sf RLCM}, \citet{hoff2005subset}, {\sf HC}, Bayesian {\sf LCA}), the percent being ranked the first or the second in terms of the mean {\sf aRI} averaged across $R=60$ replications (Section 4.2.1 in Main Paper). Each histogram is produced for the $1,920$ combinations of parameters investigated in Section 4.1 in Main Paper.}
\label{fig:hist_first_second}
\end{figure}


\begin{figure}[]
\captionsetup{width=0.8\linewidth}
\centering
\addtocounter{figure}{1} 
\includegraphics[width=0.9\linewidth]{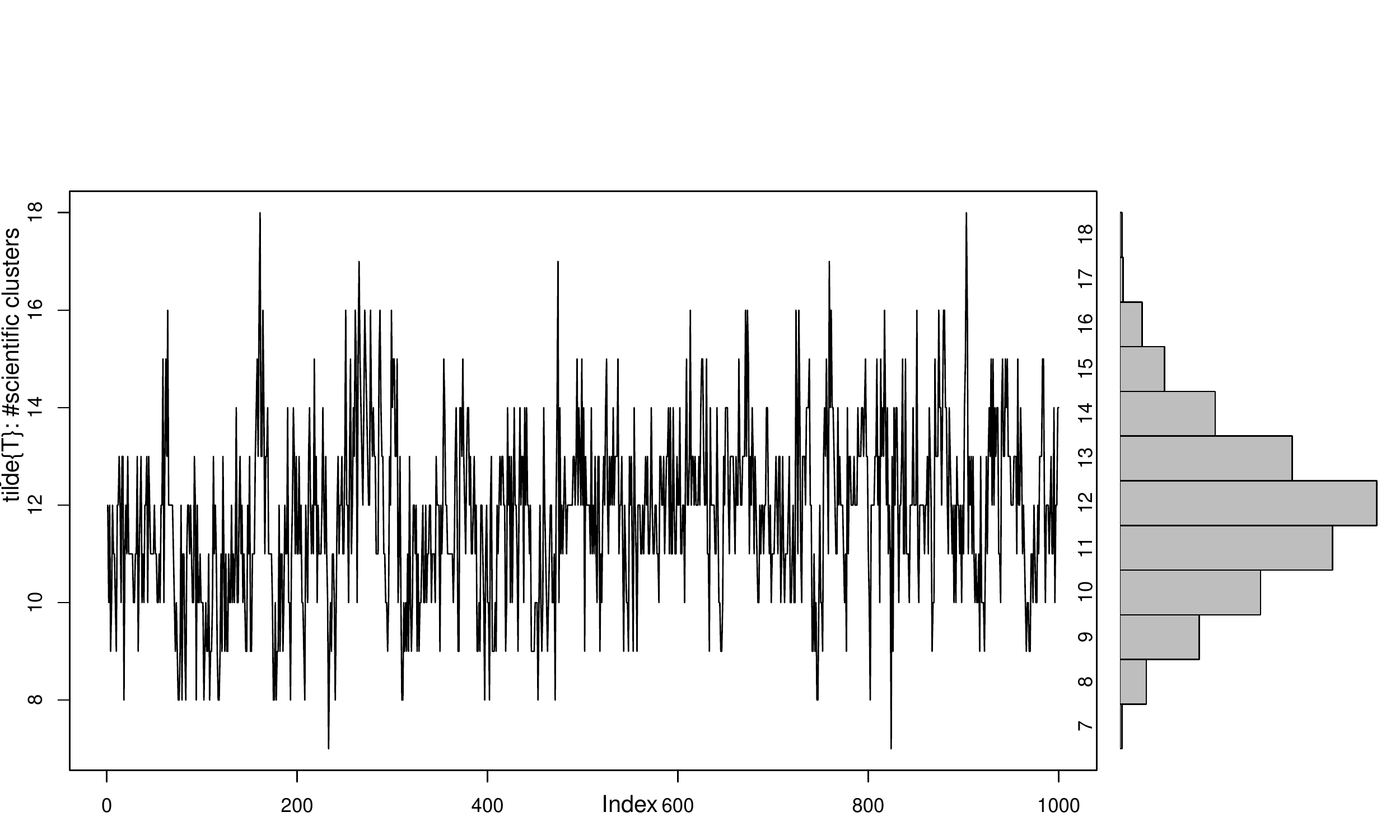}
\label{fig:truth}
\addtocounter{figure}{-1} 
\caption{MCMC samples of the number of scientific clusters $(\tilde{C})$ with its marginal posterior on the right margin.}
\label{fig:posterior_cluster_number}
\end{figure}

\begin{figure}[htp]
\captionsetup{width=0.8\linewidth}
\centering
\addtocounter{figure}{1} 
\includegraphics[width=0.9\linewidth]{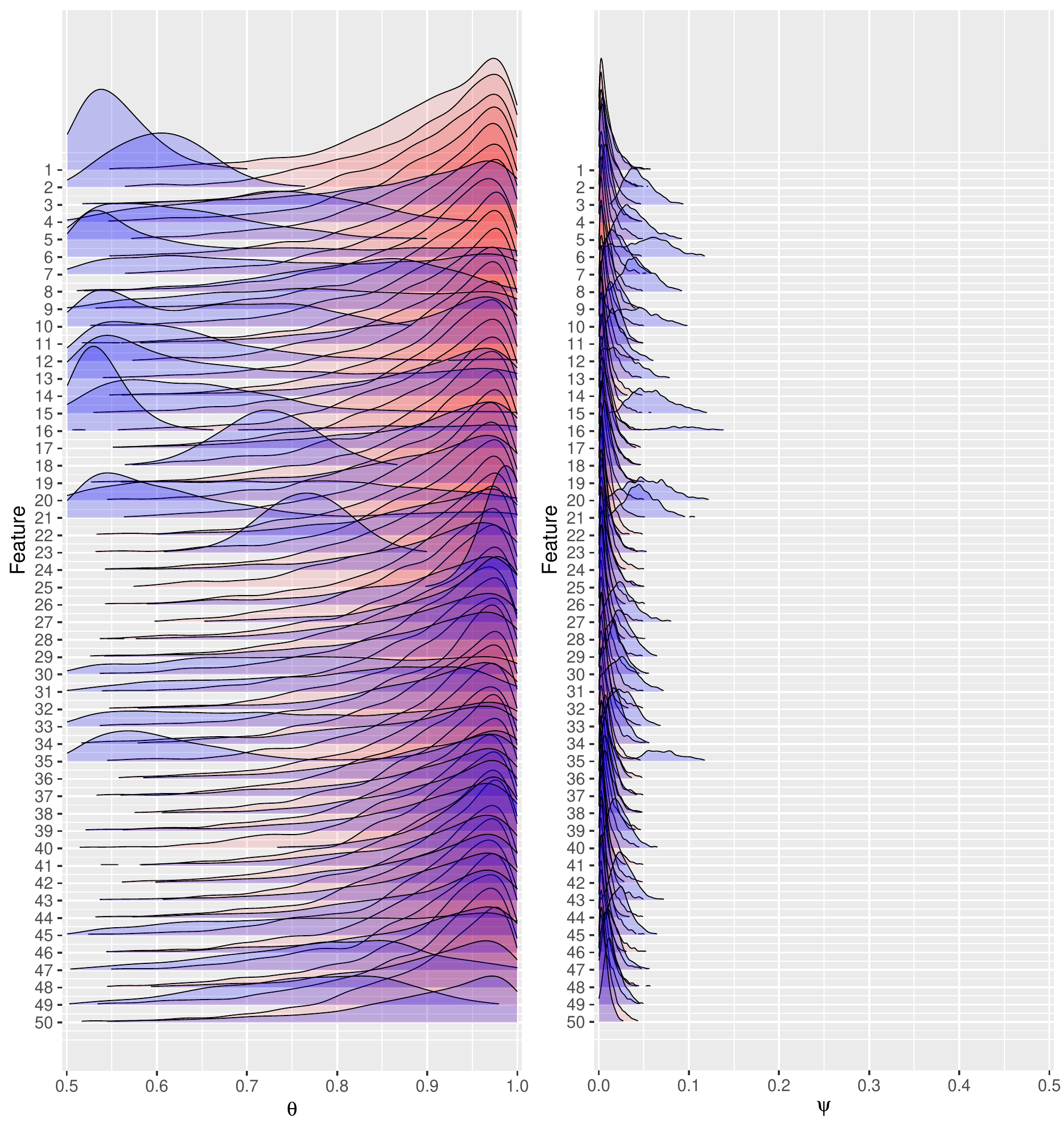}
\label{fig:truth}
\addtocounter{figure}{-1} 
\caption{Prior vs posterior for all true positive rates $\{\theta_\ell\}$ (left) and false positive rates $\{\psi_\ell\}$ (right).}
\label{fig:positive_rate_ridgelines}
\end{figure}

\begin{figure}[htp]
\captionsetup{width=0.95\linewidth}
\centering
\addtocounter{figure}{1} 
\includegraphics[width=0.95\linewidth]{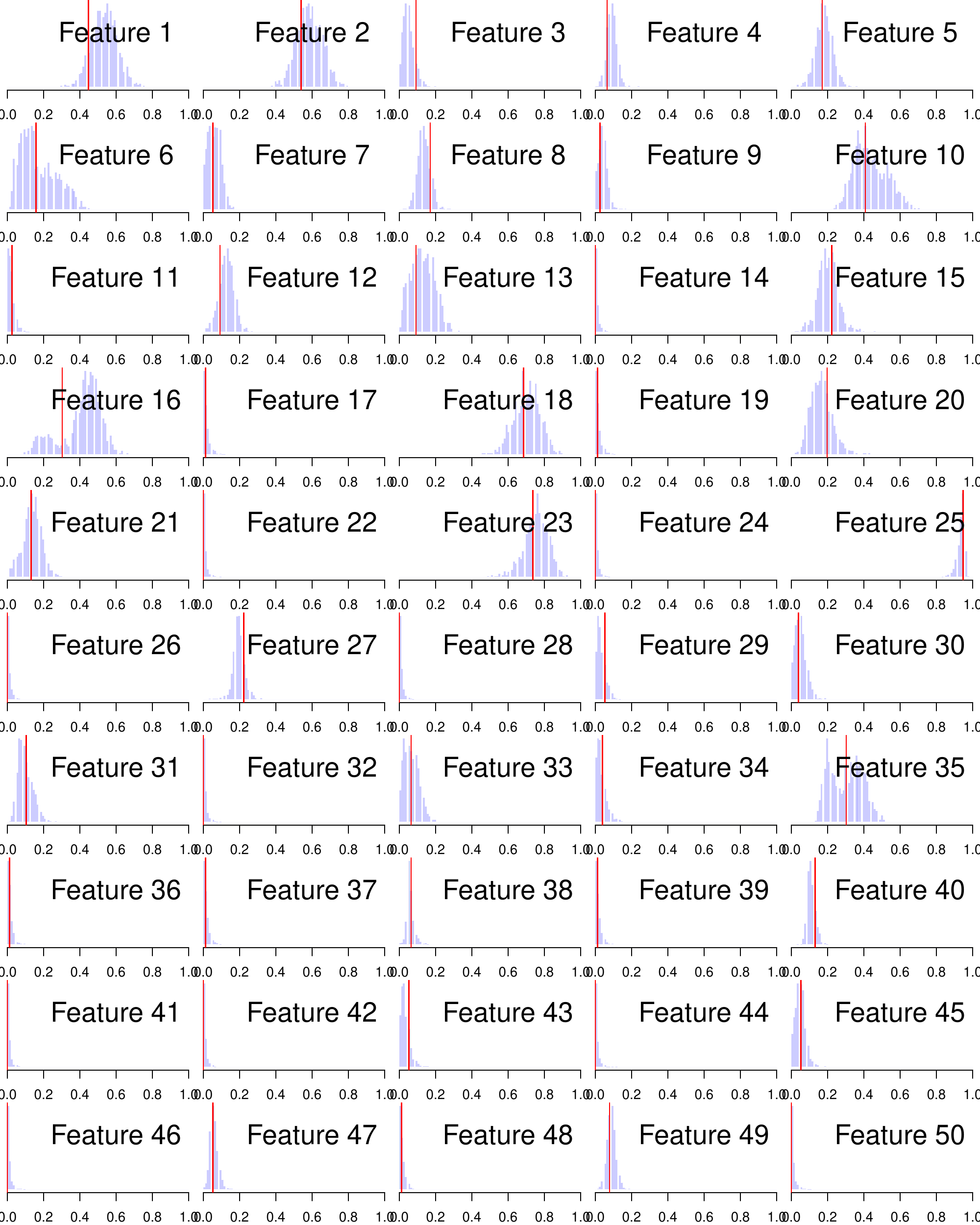}
\label{fig:truth}
\addtocounter{figure}{-1} 
\caption{Observed marginal positive rate (solid vertical line) plotted against the posterior predictive distributions for $L=50$ landmarks in Example 1.}
\label{fig:ppd_mean}
\end{figure}

\begin{figure}[htp]
\captionsetup{width=\linewidth}
\centering
\addtocounter{figure}{1} 
\includegraphics[width=\linewidth]{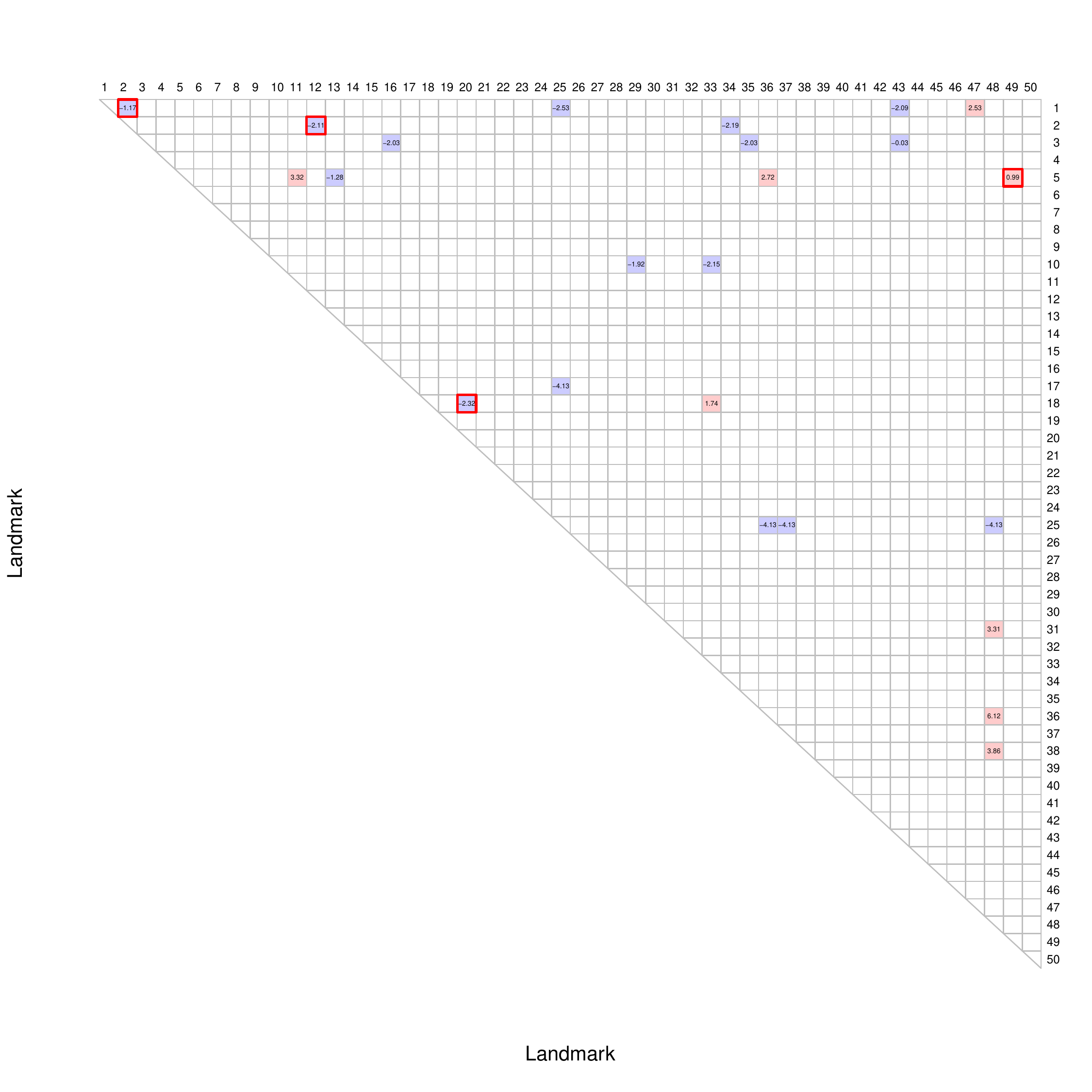}
\label{fig:truth}
\addtocounter{figure}{-1} 
\caption{Significant deviations of model predicted log odds ratios (LOR) from the observed LOR. A blank cell indicates a good model prediction for the observed pairwise LOR ($|\textsf{SLORD}| < 2$); A red (blue) cell indicates model under- (over-) fitting $\textsf{SLORD}>2 (<-2)$, where standardized LOR difference (SLORD) is defined as the observed LOR for a pair of landmarks minus the mean LOR for the predictive distribution value divided by the standard deviation of the LOR predictive distribution. A red box indicate that the pair of landmarks have cell counts in the 2 by 2 observed marginal table all greater than $5$.}
\label{fig:ppd_log_odds_ratio}
\end{figure}

\end{document}